\documentclass[journal,twoside,twocolumn]{ieeecolor} 
\usepackage{tmi}
\usepackage{cite}
\usepackage{amsmath,amssymb,amsfonts}
\usepackage{algorithmic}
\usepackage{graphicx}
\usepackage{textcomp}

\usepackage{subfigure}
\usepackage{bm}
\usepackage{adjustbox}
\usepackage[justification=centering]{caption}
\usepackage{hyperref}
\hypersetup{
    colorlinks=true,
    linkcolor=blue,
    filecolor=magenta,      
    urlcolor=cyan,
    }

\def\r{\mathbf{r}}

\def\H{\mathcal{H}}
\def\g{\mathbf{g}}
\def\L2{\mathbb{L}_2}

\def\th{\mathrm{th}}
\def\e{\mathcal{E}}

\begin{document}
\bstctlcite{IEEEexample:BSTcontrol}
\title{Joint regional uptake quantification of thorium-227 and radium-223 using a multiple-energy-window projection-domain quantitative SPECT method}

\author{
Zekun Li, Nadia Benabdallah, 
Richard Laforest, Richard L. Wahl, 
Daniel L. J. Thorek, Abhinav K. Jha$^*$, \textit{Senior Member, IEEE}
\thanks{The manuscript was submitted on 28 February 2023 for review and resubmitted on 2 January 2024 after revision. 
This work was supported by grants R01-EB031962, R01-EB031051, and R56 EB028287 awarded by the National Institute of Biomedical Imaging and Bioengineering and the NSF CAREER award 2239707.}
\thanks{Z. Li is with the
Department of Biomedical Engineering, Washington University, St. Louis, MO 63130 USA.}
\thanks{N. Benabdallah, 
R Laforest, and R. L. Wahl are
with the Mallinckrodt Institute of Radiology, Washington University, St. Louis, MO 63110, USA}
\thanks{D. L. J. Thorek is with the Department of Biomedical Engineering, Mallinckrodt Institute of Radiology, and Program in Quantitative Molecular Therapeutics, Washington University, St. Louis, MO, 63130 USA }
\thanks{*A. K. Jha is with the Department of Biomedical Engineering and Mallinckrodt Institute of Radiology, Washington University, St. Louis, MO 63130 USA (e-mail: a.jha@wustl.edu).}
}

\maketitle
\begin{abstract}
Thorium-227 ($\bm{\mathrm{^{227}Th}}$)-based $\alpha$-particle radiopharmaceutical therapies ($\bm{\alpha}$-RPTs) are currently being investigated in several clinical and pre-clinical studies. After administration, $\bm{\mathrm{^{227}Th}}$ decays to $\bm{\mathrm{^{223}Ra}}$, another $\bm{\alpha}$-particle-emitting isotope, which redistributes within the patient. Reliable dose quantification of both $\bm{\mathrm{^{227}Th}}$ and $\bm{\mathrm{^{223}Ra}}$ is clinically important, and SPECT may perform this quantification as these isotopes also emit X- and $\bm{\gamma}$-ray photons. However, reliable quantification is challenging for several reasons: the orders-of-magnitude lower activity compared to conventional SPECT, resulting in a very low number of detected counts, the presence of multiple photopeaks, substantial overlap in the emission spectra of these isotopes, and the image-degrading effects in SPECT. To address these issues, we propose a multiple-energy-window projection-domain quantification (MEW-PDQ) method that jointly estimates the regional activity uptake of both $\bm{\mathrm{^{227}Th}}$ and $\bm{\mathrm{^{223}Ra}}$ directly using the SPECT projection data from multiple energy windows. We evaluated the method with realistic simulation studies conducted with anthropomorphic digital phantoms, including a virtual imaging trial, in the context of imaging patients with bone metastases of prostate cancer who were treated with $\bm{\mathrm{^{227}Th}}$-based $\bm{\alpha}$-RPTs. The proposed method yielded reliable (accurate and precise) regional uptake estimates of both isotopes and outperformed state-of-the-art methods across different lesion sizes and contrasts, as well as in the virtual imaging trial. This reliable performance was also observed with moderate levels of intra-regional heterogeneous uptake as well as when there were moderate inaccuracies in the definitions of the support of various regions. Additionally, we demonstrated the effectiveness of using multiple energy windows and the variance of the estimated uptake using the proposed method approached the Cram\'er-Rao-lower-bound-defined theoretical limit. These results provide strong evidence in support of this method for reliable uptake quantification in $\bm{\mathrm{^{227}Th}}$-based $\bm{\alpha}$-RPTs.
\end{abstract}


\section{Introduction}
\label{sec:introduction}
Thorium-227 ($\mathrm{^{227}Th}$) conjugates are an emerging class of $\alpha$-particle radiopharmaceutical therapies ($\alpha$-RPTs) and are being investigated in several clinical and preclinical studies~\cite{baranowska2020alpha, frantellizzi2020targeted, hagemann2020advances}.
The $\alpha$-particles emitted from $\mathrm{^{227}Th}$ have a short emission range and a high linear energy transfer; thus, they can effectively ablate lesions while minimizing damage to surrounding normal tissues~\cite{tafreshi2019development,pouget2011clinical}.
However, since $\mathrm{^{227}Th}$ and its daughters distribute throughout the patient, including in radio-sensitive vital organs, it is important to determine the absorbed dose of these isotopes in the lesion and organs of the patient~\cite{larsson2020feasibility}.
This dose quantification can help adapt treatment regimens, monitor adverse events, predict therapy outcomes, and thus personalize RPTs~\cite{brans2007clinical,garin2021personalised,siegel2003red,sgouros2020dosimetry}.

The $\mathrm{^{227}Th}$ and some of its daughters also emit X- and $\gamma$-ray photons, which can be detected by a $\gamma$-camera. Thus single-photon emission computed tomography (SPECT) could provide a mechanism to quantify the activity uptake in organs and lesions in $\mathrm{^{227}Th}$-based $\alpha$-RPTs, from which the absorbed dose could be quantified. However, this quantification task faces several challenges. First, the decay chain of $\mathrm{^{227}Th}$ is complex. The first daughter of $\mathrm{^{227}Th}$ is radium-223 ($\mathrm{^{223}Ra}$), another $\alpha$-particle, as well as X- and $\gamma$-ray photon emitter, with a half-life comparable to that of $\mathrm{^{227}Th}$. $\mathrm{^{223}Ra}$ can disassociate from its original sites of decay and form a biodistribution that is independent of $\mathrm{^{227}Th}$~\cite{murray2020quantitative}.
While the subsequent daughters of $\mathrm{^{223}Ra}$ decay rapidly, the long-lived $\mathrm{^{227}Th}$ and $\mathrm{^{223}Ra}$ lead to two independent $\alpha$-emitting isotope distributions within the patient. Thus, quantifying the absorbed dose in the patient in $\mathrm{^{227}Th}$-based $\alpha$-RPTs requires estimation of the activity uptake of both $\mathrm{^{227}Th}$ and $\mathrm{^{223}Ra}$ in various regions. Second, the X- and $\gamma$-ray spectra of the two isotopes significantly overlap, resulting in substantial crosstalk contamination. This overlap presents a challenge in using conventional single-isotope quantitative SPECT methods to accurately quantify each isotope individually. Next, SPECT suffers from multiple image-degrading processes including attenuation, scatter, and imperfections with the collimator-detector response. Finally, but very importantly, similar to other $\alpha$-RPTs, the administered activity in $\mathrm{^{227}Th}$-based $\alpha$-RPT is usually two-to-three orders lower than in conventional SPECT procedures, leading to an extremely low number of detected photons.

As mentioned above, a key reason to perform quantitative SPECT in $\mathrm{^{227}Th}$-based $\alpha$-RPTs is to perform organ and lesion-level dosimetry~\cite{sgouros2020dosimetry,pandit2021dosimetry}. For this, the input to the dosimetry technique is the regional uptake in the organs and lesions. Thus, our study focuses on the task of quantifying the regional activity uptake of $\mathrm{^{227}Th}$ and $\mathrm{^{223}Ra}$ in the organs and lesions of patients undergoing $\mathrm{^{227}Th}$-based $\alpha$-RPT. A conventional framework for such quantification is to first reconstruct the activity distribution across a voxelized grid, define volumes of interest (VOIs) on that grid corresponding to regions where the activity uptake needs to be estimated, and finally, compute the average activity across all voxels within the VOI. 
We refer to quantification methods within this conventional voxelized image reconstruction-based quantification (RBQ) framework as conventional RBQ methods in the rest of the article. For these conventional RBQ methods, dual-isotope SPECT reconstruction approaches provide a mechanism to jointly reconstruct the images of $\mathrm{^{227}Th}$ and $\mathrm{^{223}Ra}$ distributions.

Existing dual-isotope SPECT reconstruction approaches can be divided into two categories. The first category compensates for the crosstalk contamination between the two isotopes prior to performing reconstruction~\cite{ichihara1993compton,moore1995simultaneous,links1996vector,feng2002accurate}.
A widely used approach in this category is triple energy window (TEW)-based crosstalk compensation, which assumes that crosstalk contamination arises mainly from scattered photons~\cite{ichihara1993compton}.
However, in the joint $\mathrm{^{227}Th}$ and $\mathrm{^{223}Ra}$ quantification, due to the highly overlapped emission spectra, crosstalk contamination is also caused by primary photons, which is challenging to compensate with the TEW-based method. In addition, the TEW method typically obtains noisy crosstalk estimates from narrow energy windows positioned on either side of the photopeak window. Compensation using these noisy crosstalk estimates leads to amplified image noise~\cite{du2007model,king1997investigation}.
The second category of methods performs model-based crosstalk compensation during reconstruction~\cite{kadrmas1999simultaneous,wang2001model,song2004validation,du2007model,du2009quantitative,de2002simultaneous,ouyang2006fast}. 
These methods can be designed to model crosstalk from both primary and scattered photons, and they have recently been studied for jointly quantifying $\mathrm{^{227}Th}$ and $\mathrm{^{223}Ra}$ activity uptake~\cite{ghaly2019quantitative}.
However, at low counts, these conventional RBQ methods have been observed to yield limited accuracy and precision on the task of quantifying regional activity uptake values, with bias and standard deviation values between 10-35\% and 12-30\%, respectively, even with fine-tuned reconstruction protocols quantifying uptake of a single isotope~\cite{ghaly2019quantitative,li2022projection,benabdallah2019223,yue2016f,gustafsson2020feasibility}.
Thus, there is an important need for new approaches on the task of jointly quantifying regional uptake of $\mathrm{^{227}Th}$ and $\mathrm{^{223}Ra}$.

The major challenge in conventional RBQ methods is that the regional uptake estimation is performed from reconstructed SPECT images. Reconstructing the images requires estimating a large number of voxel values from the projection data, an ill-posed problem that becomes even more challenging when the number of detected counts is small. Further, the image reconstruction process is also subject to information loss~\cite{barrett2013foundations,beaudry2011intuitive}. 
In this context, we recognize that reconstructing the images over a voxel basis is only an intermediate step in estimating the regional uptake. Further, the number of VOIs over which the uptake is required to be estimated is far fewer than the number of voxels. Thus, directly quantifying the mean uptake in VOIs from the projection data is a less ill-posed problem and can avoid information loss in the reconstruction step. Given these considerations, building upon ideas proposed by Carson~\cite{carson1986maximum}, we developed a projection-domain low-count quantitative SPECT (LC-QSPECT) method that estimates activity uptake in the VOIs from projection data~\cite{li2022projection}. 
The method was validated for $\mathrm{^{223}Ra}$-based $\alpha$-RPTs, and observed to yield accurate and precise regional uptake estimates and outperform conventional RBQ methods~\cite{li2022projection}. 
However, the LC-QSPECT method can only estimate the regional uptake of a single isotope from projections in the photopeak energy window. Further, the method cannot compensate for the significant crosstalk contamination of $\mathrm{^{227}Th}$ and $\mathrm{^{223}Ra}$. Our studies, as presented in the Supplementary material and in Li et al.~\cite{liGitHub}, show that applying the LC-QSPECT method to quantify the regional uptake of patients undergoing $\mathrm{^{227}Th}$-based $\alpha$-RPT leads to highly erroneous estimates.

To address the above-mentioned challenges, we recognize that both $\mathrm{^{227}Th}$ and $\mathrm{^{223}Ra}$ have distinct emission spectra comprising X- and $\gamma$-ray emissions over multiple photopeak energies. This results in a unique system response across various energy windows when imaging each of these two isotopes with SPECT. Consequently, we can use projection data across these multiple energy windows while accounting for the crosstalk contamination to directly and jointly estimate the regional uptake of both $\mathrm{^{227}Th}$ and $\mathrm{^{223}Ra}$. This approach of accounting for crosstalk contamination in joint dual-isotope estimation aligns with model-based crosstalk contamination compensation approaches used in the conventional RBQ methods~\cite{du2007model,li2022projection,benabdallah2019223,yue2016f,gustafsson2020feasibility}.
Further, using photons from multiple energy windows corresponding to these different photopeaks provides a mechanism to increase the photon counts and thus improve the effective system sensitivity~\cite{benabdallah2019223}.
Recent studies have shown that using measurements from multiple energy windows can help improve quantification performance~\cite{rahman2020list,chun2019algorithms}.
Integrating these ideas, we propose a multiple-energy-window projection-domain quantification (MEW-PDQ) method that jointly estimates the regional activity uptake of $\mathrm{^{227}Th}$ and $\mathrm{^{223}Ra}$ directly from low-count SPECT projections acquired over multiple energy windows. Preliminary versions of the work described in this paper have been presented previously~\cite{li2022multiple,schramm2021proceedings}.

\section{Theory}
Consider a SPECT system that images a radioisotope distribution of both $\mathrm{^{227}Th}$ and $\mathrm{^{223}Ra}$. Both $\mathrm{^{227}Th}$ and $\mathrm{^{223}Ra}$ emit photons at multiple energies. Denote the overall$\ $photon source distribution as $\Xi\left( \bm{r}\mathcal{,E} \right)$, which describes the mean emission rate of $\gamma$-ray and X-ray photons with energy $\mathcal{E}$ at the 3-dimensional (3D) position $\bm{r}$. Denote the measured projection data by a $M$-dimensional vector $\bm{g}$, containing projections in multiple energy windows. Denote the number of projection bins with different spatial positions by $M_{s}$, and the number of energy windows by $M_{e}$. Thus, $M = M_{s} \times M_{e}$. 

Assume that the photon source distribution being imaged and the projection data lie in the Hilbert space of square-integrable functions, denoted by $\mathbb{L}_{2}\left( \mathbb{R}^{4} \right)$, and the Hilbert space of Euclidean vectors, denoted by $\mathbb{E}^{M}$, respectively. Then, the SPECT system imaging the photon source distribution, denoted by the operator $\mathcal{H}$, can be described as an integral transform from $\mathbb{L}_{2}\left( R^{4} \right)$ to $\mathbb{E}^{M}$. Denote the kernel of the $\mathcal{H}$ operator by $h_{m}\left( \bm{r}\mathcal{,E} \right)$, which characterizes the system response of the $m^{th}$ element of projection $\bm{g}$ to a photon emitted at position $\bm{r}$ with energy $\mathcal{E}$. We note here that the emitted photon may be either a $\gamma$-ray or an X-ray photon, and the system response models the probability of detecting any counts due to the emitted photon. This includes the probability of detecting characteristic X-ray photons from the collimator caused by the emitted photon. 

In SPECT with $\alpha$-RPTs, due to the very low count levels, the stray-radiation-related noise due to photons emitted from regions other than the patient forms a substantial portion of the measured counts. We model this noise as Poisson distributed, with the same mean for all projection bins in a given energy window but different means for different energy windows. Let $\bm{\Psi}$ be an $M$-dimensional vector that denotes the mean stray-radiation-related noise in the $M$ projection bins. The imaging system equation is given by 
\begin{equation}
\label{eq:system_eq}
\begin{split}
    \bm{g} = \H \bm{\Xi} + \bm{\Psi} + \bm{n},
\end{split}
\end{equation}
where $\bm{n}$, an $M$-dimensional vector, denotes the entire noise in the imaging system. Overall, the data $\bm{g}$ is Poisson distributed with mean of $\mathcal{H}\bm{\Xi} + \bm{\Psi}$. The $m^{th}$ element of this vector is given by: 
\begin{equation}
\label{eq:gm}
    g_m = \int \int h_m(\bm{r},\mathcal{E}) \Xi(\bm{r},\mathcal{E})~d^3 r~d\mathcal{E} + \psi_m + n_m.
\end{equation}

The term $\Xi\left( \bm{r}\mathcal{,E} \right)$ comprises photons emitted by both $\mathrm{^{227}Th}$ and $\mathrm{^{223}Ra}$. Denote the mean emission rate of$\ $photons from $\mathrm{^{227}Th}$ and $\mathrm{^{223}Ra}$ with energy $\mathcal{E}$ at position $\bm{r}$ by $\Xi^{Th}\left( \bm{r}\mathcal{,E} \right)$ and $\Xi^{Ra}\left( \bm{r}\mathcal{,E} \right)$, respectively. $\Xi^{Ra}\left( \bm{r}\mathcal{,E} \right)$ includes the photons emitted by both $\mathrm{^{223}Ra}$ and its daughter isotopes. Then
\begin{equation}
\label{Eq:sp_comp}
    \Xi(\bm{r},\mathcal{E}) = \Xi^{Th}(\bm{r},\mathcal{E}) + \Xi^{Ra}(\bm{r},\mathcal{E}).
\end{equation}
Since the emission rates of photons with different energies are independent of the spatial location, we can decompose $\Xi^{Th}\left( \bm{r}\mathcal{,E} \right)$ and $\Xi^{Ra}\left( \bm{r}\mathcal{,E} \right)$ as
\begin{subequations}
\label{Eq:ind_sp_comp}
\begin{align}
    \Xi^{Th}(\bm{r},\mathcal{E})&=  f^{Th}(\bm{r}) \zeta^{Th}(\mathcal{E}), \\
    \Xi^{Ra}(\bm{r},\mathcal{E})&=  f^{Ra}(\bm{r}) \zeta^{Ra}(\mathcal{E}),
\end{align}
\end{subequations}
where $\zeta^{Th}\left( \mathcal{E} \right)$ and $\zeta^{Ra}\left( \mathcal{E} \right)$ describe the mean emission rates of photons with energy $\mathcal{E}$ from a unit activity of $\mathrm{^{227}Th}$ and $\mathrm{^{223}Ra}$, respectively, and $f^{Th}\left( \bm{r} \right)$ and $f^{Ra}\left( \bm{r} \right)$ denote the spatial distribution of these isotopes.
Inserting Eqs.~\eqref{Eq:sp_comp} and~\eqref{Eq:ind_sp_comp} into Eq.~\eqref{eq:gm}, we obtain
\begin{equation}
\label{eq:gm_explicit}
\begin{split}
    &g_m = \int  f^{Th}(\r) \int h_m(\r,\e) \zeta^{Th}(\e)~d\e ~d^3 r  \\
    &+\int f^{Ra}(\r) \int h_m(\r,\e) \zeta^{Ra}(\e)~d\e ~d^3r +\psi_m + n_m.
\end{split}
\end{equation}
We define
\begin{subequations}
\begin{align}
    h^{Th}_m(\bm{r}) &= \int h_m(\bm{r},\e) \zeta^{Th}(\e)~d\e,\\
    h^{Ra}_m(\bm{r}) &= \int h_m(\bm{r},\e) \zeta^{Ra}(\e)~d\e.
\end{align}
\end{subequations}
The terms $h_{m}^{Th}\left( \bm{r} \right)$ and $h_{m}^{Ra}\left( \bm{r} \right)$ can be considered as kernels of two isotope-dependent system operators, describing the system response of the $m^{th}$ element of $\bm{g}$ to a unit activity uptake of $\mathrm{^{227}Th}$ and $\mathrm{^{223}Ra}$ at position $\bm{r}$, respectively. 
Eq.~\eqref{eq:gm_explicit} can be written as
\begin{equation}
\label{eq:gm_exp_simp}
\begin{split}
        g_m &= \int \left\{ h^{Th}_m(\r)f^{Th}(\r) 
        +  h^{Ra}_m(\r)f^{Ra}(\r) \right\} ~d^{3}r +\psi_m + n_m.
\end{split}
\end{equation}

In Eq.~\eqref{eq:gm_exp_simp}, we defined the imaging equation in terms of the spatial distribution of $f^{Th}\left( \bm{r} \right)$ and $f^{Ra}\left( \bm{r} \right)$. We recognize that our objective is to estimate the absolute regional activity uptake of both isotopes within a set of VOIs. Mathematically, we first define a 3D VOI function $\phi_{k}^{VOI}\left( \bm{r} \right)$, where
\begin{equation}
\label{eq:dif_roi}
\phi^{VOI}_{k}(\bm{r}) =\left\{
        \begin{array}{cc}
             1,&{\mathrm{if~}\bm{r}\mathrm{~lies~within~the~k^{th}~VOI.}} \\
             0,&  {\mathrm{otherwise}}.
        \end{array}
\right.
\end{equation}
Denote $\bm{\lambda}^{Th}$ and $\bm{\lambda}^{Ra}$ as $K$-dimensional vectors of regional uptake of $\mathrm{^{227}Th}$ and $\mathrm{^{223}Ra}$, respectively. $\lambda_{k}^{Th}$ and $\lambda_{k}^{Ra}$ are given by 
\begin{subequations}
\begin{align}
\label{eq:th_basis}
\lambda^{Th}_k &= \frac{\int d^3 r~f^{Th}(\r) \phi_k^{VOI}(\r)}{\int d^3 r~\phi_k^{VOI}(\r)}, \\
\label{eq:ra_basis}
\lambda^{Ra}_k &= \frac{\int d^3 r~f^{Ra}(\r) \phi_k^{VOI}(\r)}{\int d^3 r~\phi_k^{VOI}(\r)},
\end{align}
\end{subequations}
respectively. Our objective is to estimate $\bm{\lambda}^{Th}$ and $\bm{\lambda}^{Ra}$.

Toward this goal, we directly represent the activity uptake distributions $f^{Th}(\bm{r})$ and $f^{Ra}(\bm{r})$ in terms of the VOI-basis functions as given by Eq.~\eqref{eq:dif_roi}. The activity distributions of $\mathrm{^{227}Th}$ and $\mathrm{^{223}Ra}$ are then represented in terms of these basis functions as
\begin{subequations}
\label{eq:voi_def}
\begin{align}
f^{Th}_{VOI} (\r ) &= \sum_{k=1}^K \lambda^{Th}_k \phi_k^{VOI}(\r),\\
f^{Ra}_{VOI} (\r ) &= \sum_{k=1}^K \lambda^{Ra}_k \phi_k^{VOI}(\r),
\end{align}
\end{subequations}
respectively.
Inserting the VOI-based representations for $f^{Th}(\r)$ and $f^{Ra}(\r)$ in Eq.~\eqref{eq:gm_exp_simp} yields the following expression for the $m^{th}$ element of the vector $\g$:
\begin{equation}
\label{eq:gm_voi}
    \begin{split}
    g_m &= \sum_{k=1}^{K} \lambda^{Th}_{k}\int h^{Th}_m(\r)\phi_{k}^{VOI}(\r)~d^{3}r \\
        &+ \sum_{k=1}^{K} \lambda^{Ra}_{k}\int h^{Ra}_m(\r)\phi_{k}^{VOI}(\r)~d^{3}r + \psi_m +n_m.
    \end{split}
\end{equation}
To simplify the notation, we define $\bm{H}^{Th}$ and $\bm{H}^{Ra}$ as two $M \times K$ dimensional system matrices with elements given by 
\begin{subequations}
\begin{align}
    H^{Th}_{mk} &= \int d^3 r ~h^{Th}_m(\r) \phi_k^{VOI} (\r),\label{eq:sys_mats1}\\
    H^{Ra}_{mk} &= \int d^3 r ~h^{Ra}_m(\r) \phi_k^{VOI} (\r),\label{eq:sys_mats2}
\end{align}
\end{subequations}
respectively. 
Then, Eq.~\eqref{eq:gm_voi} can be written as 
\begin{equation}
    g_m = \sum_{k=1}^{K} H_{mk}^{Th} \lambda_k^{Th} + \sum_{k=1}^{K} H_{mk}^{Ra}\lambda_k^{Ra} + \psi_m + n_m.
\end{equation}
This can then be written in vector form as
\begin{equation}
\label{eq:vec_comp}
\begin{split}
    \bm{g} &= \bm{H}^{Th} \bm{\lambda}^{Th} + \bm{H}^{Ra} \bm{\lambda}^{Ra} + \bm{\Psi} + \bm{n}.
\end{split}
\end{equation}
To make the notation more concise, we define the following terms:
\begin{gather}
    \bm{H} = 
    \begin{bmatrix}
    \bm{H}^{Th}&\bm{H}^{Ra}
    \end{bmatrix},~
    \bm{\lambda} = 
    \begin{bmatrix}
    \bm{\lambda}^{Th}\\
    \bm{\lambda}^{Ra}
    \end{bmatrix}.
\end{gather}
Then, Eq.~\eqref{eq:vec_comp} can be written as
\begin{equation}
\label{eq:vec_simp}
    \bm{g} = \bm{H} \bm{\lambda} + \bm{\Psi} + \bm{n}.
\end{equation}

Having derived this mathematical formalism to estimate $\bm{\lambda}$, we derive a maximum-likelihood (ML) expectation maximization (EM) approach.
Denote the probability of a discrete random variable $x$ as $\mathrm{Pr}(x)$. Then the probability of the measured projection data is given by
\begin{equation}
    \Pr(\g|\bm{\lambda}) = \prod_{m=1}^M \Pr(g_m |\bm{\lambda}), 
\end{equation}
where we have used the fact that the measured data across the different bins and energy windows are independent. Now, from Eq.~\eqref{eq:vec_simp}, the measured data $g_m$ is Poisson distributed with mean $(\bm{H \lambda})_m + \psi_m$. Thus
\begin{equation}
\begin{split}
    \Pr(\g|\bm{\lambda})  &= \prod_{m=1}^{M}\exp[-(\bm{H}^{Th} \bm{\lambda}^{Th} + \bm{H}^{Ra} \bm{\lambda}^{Ra} )_m - \psi_m] \\ 
    &\times \frac{[(\bm{H}^{Th} \bm{\lambda}^{Th} + \bm{H}^{Ra} \bm{\lambda}^{Ra})_m+\psi_m]^{g_m}}{{g_m}!}.
\end{split}
    \label{eq:poisson}
\end{equation}
This gives the likelihood of the measured data $\g$.
To estimate $\bm{\lambda}$, we maximize the logarithm of this likelihood of $\bm{\lambda}$ given $\g$.
Briefly, we differentiate the log-likelihood with respect to the elements of $\bm{\lambda}$ and equate that to $0$ to find the point at which the log-likelihood is maximized. This yields the following coupled equations to estimate the regional activity uptake of $\mathrm{^{227}Th}$ and $\mathrm{^{223}Ra}$ in the $k^{th}$ VOI:
\begin{subequations}
\begin{align}
    \widehat{\lambda^{Th}_{k}}^{(t+1)} &= \frac{\widehat{\lambda^{Th}_{k}}^{(t)}}{\sum\limits_{m=1}^{M} H^{Th}_{mk}}
   \sum_{m=1}^{M}
   \frac{g_{m}}{[\bm{H\hat{\lambda}}^{(t)}]_{m}+\psi_m} H^{Th}_{mk}, \label{eq:MLEM}\\
    \widehat{\lambda^{Ra}_{k}}^{(t+1)} &= \frac{\widehat{\lambda^{Ra}_{k}}^{(t)}}{\sum\limits_{m=1}^{M} H^{Ra}_{mk}}
   \sum_{m=1}^{M}
   \frac{g_{m}}{[\bm{H\hat{\lambda}}^{(t)}]_{m}+\psi_m} H^{Ra}_{mk},\label{eq:MLEM2}
\end{align}
\end{subequations}
where $\widehat{\lambda^{Th}_{k}}^{(t)}$ and $\widehat{\lambda^{Ra}_{k}}^{(t)}$ denote the estimates of $\lambda^{Th}_k$ and $\lambda^{Ra}_k$ in the $t^{th}$ iteration.

The proposed MEW-PDQ method has several features that make it effective for performing the task of jointly estimating the regional uptake of $\mathrm{^{227}Th}$ and $\mathrm{^{223}Ra}$.
First, the method makes use of projection data from multiple energy windows for the estimation task. This not only increases the effective system sensitivity but can also enable the use of information present in data across the multiple energy windows, which could improve the theoretical limits of precisely estimating the regional uptake~\cite{chun2019algorithms}.
This improvement, as shown by our studies presented later, is significant for this joint estimation task. Second, the proposed method is less ill-posed than the conventional RBQ methods as described in Sec.~\ref{sec:introduction}. The number of VOI uptake, $K$, directly estimated by the proposed method, is much lower than the number of voxels in a reconstructed SPECT image, which is usually obtained as an intermediate step to estimate the VOI uptake values in conventional RBQ methods. Next, the method is less sensitive to partial volume effects (PVEs)~\cite{soret2007partial}, a major degrading effect in conventional image reconstruction-based quantitative SPECT. PVEs arise due to two factors, the first being the finite resolution of the system, leading to blurs in the reconstructed image, and the second being the pixelized representation of a continuous object (tissue-fraction effects). The proposed method is less affected by both these factors since we define the boundaries of VOIs from high-resolution images and directly estimate the regional uptake from SPECT projections, and since there are no voxelizations. 
Finally, the proposed method directly estimates the regional uptake from the projection data. This avoids potential errors introduced in the image reconstruction step or reconstruction-related information loss~\cite{barrett2013foundations,beaudry2011intuitive}. 

\section{Implementation and evaluation of the proposed method}
\label{sec:evaluation}
\subsection{Implementation}
\label{sec:implementation}
To implement the proposed MEW-PDQ method, we first need to obtain the elements of the system matrices corresponding to imaging $\mathrm{^{227}Th}$ and $\mathrm{^{223}Ra}$, as shown in Eqs.~\eqref{eq:sys_mats1} and~\eqref{eq:sys_mats2}, respectively.
For this purpose, definitions of the support of VOIs, as defined in Eq.~\eqref{eq:dif_roi}, are necessary. These VOI definitions can be obtained by segmenting the X-ray CT acquired in conjunction with the SPECT~\cite{da2001absolute,ritt2020quantitative}. Further, considering that patients undergoing $\alpha$-RPTs usually have scans conducted previously, such as pre-therapy PET scans, these scans could also be used in defining the VOIs. Notably, even though the proposed method does not estimate regional uptake from the reconstructed images, a reconstructed SPECT image fused with an X-ray CT image could potentially be used to define the VOIs as well. 

\begin{figure}
    \centering
    \includegraphics[width=0.45\textwidth]{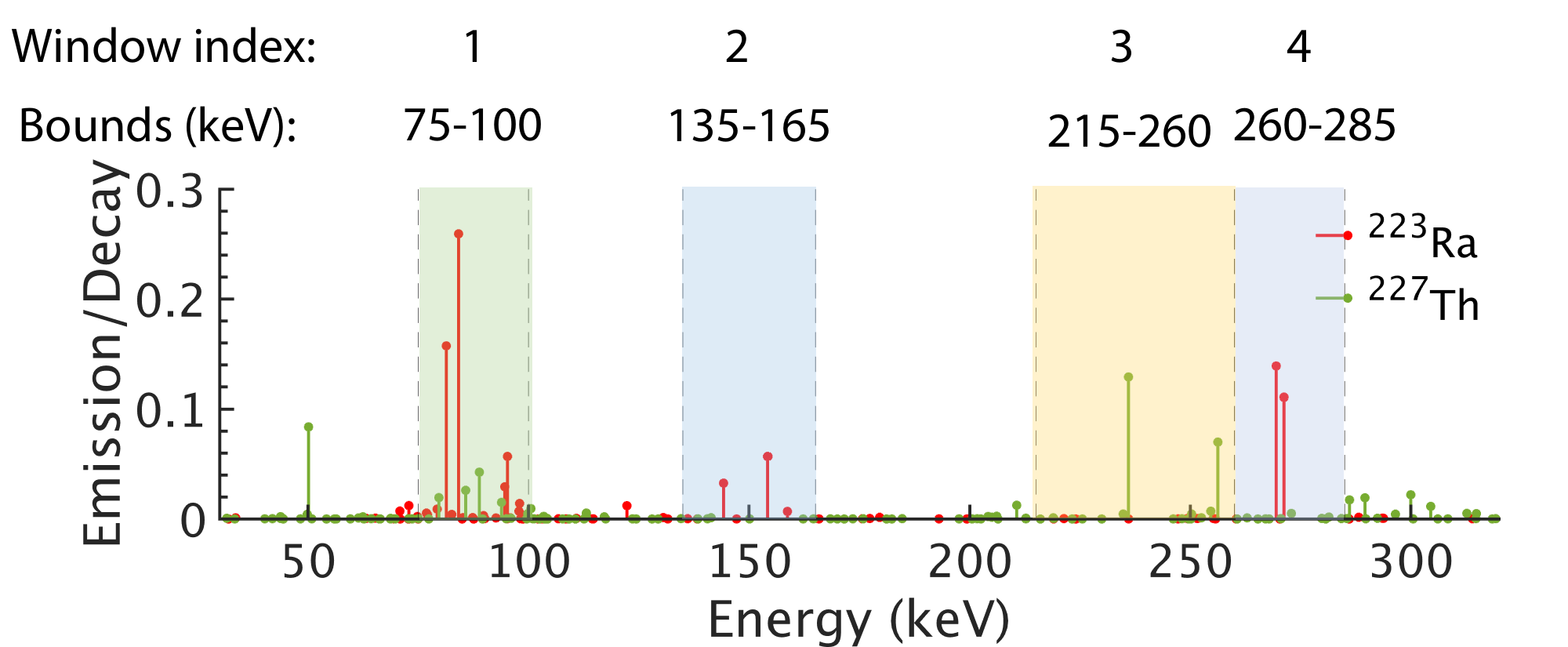}
    \caption{Bounds of different energy windows and photon emission spectra of $\mathrm{^{227}Th}$ and $\mathrm{^{223}Ra}$.}
    \label{fig:emi_spect}
\end{figure}

With the VOI definitions established, we used a Monte Carlo (MC)-based approach to obtain elements of the system matrices. Specifically, we used SIMIND, a well-validated MC-based simulation software~\cite{ljungberg1989monte,toossi2010simind,morphis2021validation}.
For each isotope, we modeled the entire emission spectrum that contained $\gamma$-ray and X-ray photons with different energies. The spectrum of \textsuperscript{223}Ra contained photons emitted by all its daughter isotopes. We assigned unit activity uptake to the VOI mask and zero uptake elsewhere. Assuming that the attenuation map of the whole patient was available, using SIMIND, we generated the projection of this activity map. For each isotope and each VOI, we simulated more than 500 million decay events to ensure a low Poisson noise in the generated projection.
As suggested in~\cite{larsson2020feasibility}, to cover the major photopeaks of both isotopes, we considered the detection of photons in four energy windows (Fig.~\ref{fig:emi_spect}).
The simulations modeled all relevant image-degrading processes in SPECT, including the attenuation, scatter, collimator response, septal penetration and scatter, characteristic X-ray from both the $\alpha$-emitting isotopes and the lead in the collimator, finite energy and position resolution of the detector, scatter of the photons in the crystal and photomultiplier tubes, as well as backscatter in the detector. The energy dependency of these image-degrading processes was also modeled. Normalizing the projection data yielded the corresponding column of the system matrix for each isotope. 

Next, we obtained the values of mean stray-radiation-related noise $\psi_m$ in Eqs.~\eqref{eq:MLEM} and~\eqref{eq:MLEM2}. The mean of this noise in energy window 3 (Fig.~\ref{fig:emi_spect}) was determined from a planar blank scan acquired on an actual GE Discovery 670 SPECT/CT system with high energy general purpose (HEGP) collimator for 10~minutes. Averaging the counts in projection bins in the corresponding energy window of this scan yielded the mean background counts for 10~minutes, which were then scaled according to the acquisition time of patient images to estimate the mean of stray-radiation-related noise in this energy window. The mean values of the stray-radiation-related noise for the other energy windows in Fig.~\ref{fig:emi_spect} were estimated by scaling the mean of energy window 3 by the relative width of the corresponding energy window. We assumed here that the mean values of stray-radiation-related noise were proportional to the widths of the energy windows. The computed system matrix and means of stray-radiation-related noise were used in Eqs.~\eqref{eq:MLEM} and~\eqref{eq:MLEM2} to estimate the regional uptake directly from the projection data. The process to generate the system matrix and simulate the noise was validated by comparing the data generated using our simulation process to the measured data on an actual SPECT scanner. Details of this validation are provided in the Supplementary material as well as in Li et al.~\cite{liGitHub}.

\subsection{Evaluation framework}
Evaluating the performance of the MEW-PDQ method required a setup where the ground-truth regional uptake is known. For this purpose, we conducted realistic simulation-based studies, including a virtual imaging trial (VIT). These simulations were conducted in the context of imaging patients with bone metastases of prostate cancer who were treated with $\mathrm{^{227}Th}$-based $\alpha$-RPTs. We focused on lesions in the pelvis, a major site of osseous metastases, located near the prostate.

\subsubsection{Generating the patient phantom}
\label{sec:pht_gen}
We used the Extended Cardiac-Torso (XCAT) phantom to generate realistic digital 3D activity and attenuation maps of patient torsos~\cite{segars2008realistic}.
To simulate continuously distributed activities and attenuation coefficients, we considered high-resolution activity and attenuation maps, namely 512 $\times$ 512 along the axial dimension and 364 slices along the depth dimension, with voxel side length of 1.105 mm. Each patient had an identically located lesion in the pelvic bone.
Based on our analysis of clinical SPECT of patients administered $\mathrm{^{223}Ra}$ therapy, the lesion, bone, and gut are the three primary sites of uptake~\cite{li2022projection}.
The rest of the region in the patient typically has a similar and low uptake, which we refer to as the background.
Based on this observation, we assumed that the activity uptake distribution of $\mathrm{^{227}Th}$ would have a similar pattern across these four VOIs. The heights, weights, lesion sizes, and activity uptake distributions of $\mathrm{^{227}Th}$ and $\mathrm{^{223}Ra}$ of the simulated patients were specific to each experiment, as we will describe in Sec.~\ref{Sec:Exp}.

\subsubsection{Modeling the SPECT system}
\label{sec:sim_eva}
We simulated the scenario where the torso region of a patient was imaged for 30~minutes using a dual-head GE Discovery 670 SPECT/CT system equipped with NaI(TI) crystals and a HEGP collimator~\cite{benabdallah2021practical}. The simulations were performed with SIMIND.
The intrinsic spatial and energy resolution of the scintillation detector in the system were 3.9~mm and 9.8~\% at 140~keV, respectively. Projections were acquired in four energy windows, as indicated in Fig.~\ref{fig:emi_spect}, and at 60 angular positions spaced uniformly over 360$^{\circ}$. We simulated the emission spectra of both isotopes, the stray-radiation-related noise, and all relevant image-degrading processes in SPECT. In the following sections, each instance of SPECT projection generated for a simulated patient is referred to as a noise realization.

\subsubsection{Estimating the uptake using the proposed method and other compared methods}
We quantified the regional uptake from the simulated projections using the proposed method as described in Sec.~\ref{sec:implementation}.
For comparative purposes, we also quantified the uptake with two state-of-the-art conventional RBQ methods: 1) a dual-isotope ordered subset expectation maximization (DOSEM)-based RBQ method that we developed based on~\cite{du2007model,du2009quantitative}, which shares similarities with the method proposed in~\cite{ghaly2019quantitative}, and 2) the DOSEM method with geometric transfer matrix (GTM)-based PVE compensation~\cite{rousset1998correction}.
These methods were specifically developed and fine-tuned to address the joint quantification problem of $\mathrm{^{227}Th}$ and $\mathrm{^{223}Ra}$. The implementation and optimization details of the DOSEM and GTM-based methods are provided in the Supplementary material and Li et al.~\cite{liGitHub}. 

\subsubsection{Figures of merit}
\label{sec:FOM}
We evaluated the accuracy, precision, and overall error of the considered methods on the task of estimating regional uptake. 
When the goal was to quantify performance over multiple noise realizations with a single patient, normalized bias (NB), normalized standard deviation (NSD), and normalized root mean square error (NRMSE) were used to quantify the accuracy, precision, and overall error, respectively.
To evaluate the accuracy and overall error of these methods over a patient population, each patient with one or multiple noise realizations, we used ensemble NB and ensemble NRMSE. 
Finally, our results indicated that the proposed method was almost unbiased. Thus, we compared the NSD of the estimates obtained with the proposed method with the square root of the  Cram\'er-Rao lower bound (CRLB) normalized by the ground truth uptake (referred to as the CRLB-derived NSD). Note that the CRLB is the minimum variance that can be achieved by an unbiased estimator.
Details on calculating the figures of merit are provided in the Supplementary material as well as in Li et al.~\cite{liGitHub}.

\subsection{Experiments}
\label{Sec:Exp}
In this section, we describe the experiments we conducted to evaluate the performance of the MEW-PDQ method. First, we evaluated the convergence, as well as the performance with various lesion sizes and contrasts of the proposed method under conditions where the uptake within the VOI was homogeneous. We also evaluated the efficacy of using multiple energy windows. Additionally, a VIT was conducted to assess the performance of the method in a population setting. Next, we evaluated the proposed method in settings where there were varying levels of heterogeneous activity uptake within the VOIs and when there were inaccuracies in the VOI definitions.

\subsubsection{Evaluating the convergence of the MEW-PDQ method}
\label{sec:convergence}
In this experiment, the simulated patients had the same height and weight as a 50$^{th}$ percentile male adult in the US~\cite{segars2008realistic}, which we refer to as average patient size. The patient torsos contained a total of 11~MBq activity of both $\mathrm{^{227}Th}$ and $\mathrm{^{223}Ra}$, referred to as standard total uptake. The regional uptake ratios of the two isotopes in the four VOIs of the patients are shown in Table~\ref{tb:uptake}. Based on the analysis of the clinical SPECT of patients receiving $\mathrm{^{223}Ra}$ therapy, these values were chosen to simulate the differences in uptake in different VOIs. The convergence of the proposed method was evaluated for lesions with different diameter values. We simulated three patients, each had a lesion with diameters: 15 mm, 25 mm, and 35 mm, respectively. We generated one noise realization for each patient, as described in Sec.~\ref{sec:sim_eva}, then applied the MEW-PDQ method for 1500 iterations, and computed the error in the estimated lesion uptake for both isotopes after every 50 iterations for each patient.
\begin{table}[]
\centering
\caption{Activity uptake ratios of $\mathrm{^{227}Th}$ and $\mathrm{^{223}Ra}$ in VOIs, simulating the difference in regional uptake and independent biodistribution of the two isotopes.}
\begin{tabular}{ccccc}
    & Background & Bone & Gut & Lesion \\ \hline
$\mathrm{^{223}Ra}$ & 2          & 5    & 25  & 20     \\
$\mathrm{^{227}Th}$ & 12         & 30   & 100 & 300   
\end{tabular}
\label{tb:uptake}
\end{table}

\subsubsection{Evaluating performance with different lesion sizes and contrasts}
Following procedures similar to that described in Sec.~\ref{sec:convergence}, we generated five patient phantoms. Each patient had a lesion diameter ranging from 15~mm to 35~mm in 5~mm increments. Then, to evaluate the sensitivity of the method to the lesion-to-bone uptake ratio (LBUR), we generated five more patient phantoms with average patient size and standard total uptake. Each patient had a lesion with a diameter of 33.75 mm, which corresponds to the average lesion diameter reported in a study involving 29 bone lesions from 6 patients with castration-resistant prostate cancer~\cite{murray2017potential}. Each patient had a different LBUR for both isotopes, ranging from 2:1 to 6:1, while the activity uptake ratios in the background, bone, and gut regions are shown in Table~\ref{tb:uptake}. We generated 50 noise realizations for each patient.

\subsubsection{Evaluating performance with different $\mathrm{^{227}Th}$ to $\mathrm{^{223}Ra}$ uptake ratios}
In dosimetry studies, typically, the patients are imaged at multiple time points post-administration. The $\mathrm{^{227}Th}$ to $\mathrm{^{223}Ra}$ uptake ratio in the patient changes over time due to the decay of both isotopes. Thus, it is important to evaluate the performance of the methods for different $\mathrm{^{227}Th}$~to~$\mathrm{^{223}Ra}$ uptake ratio values. We simulated the scenario where a patient with average patient size and a 33.75 mm diameter lesion was initially administered with 11 MBq $\mathrm{^{227}Th}$ activity and was imaged at 40, 82, 141, 500 h after administration, leading to $\mathrm{^{227}Th}$ to $\mathrm{^{223}Ra}$ uptake ratios of 10:1, 5:1, 3:1, and 1:1, respectively. We considered both the decay of $\mathrm{^{227}Th}$ and $\mathrm{^{223}Ra}$, but the organ-specific half-life of the isotopes in the patient was not modeled. Thus, for each isotope, the ratios of uptake among different VOIs remained the same as indicated in Table~\ref{tb:uptake}. For each $\mathrm{^{227}Th}$ to $\mathrm{^{223}Ra}$ uptake ratio, we generated 50 noise realizations of the patient.

\subsubsection{Evaluating the impact of using multiple energy windows on the precision of the proposed method}
To quantitatively assess the impact of using multiple energy windows on the theoretical lower bound of the estimated regional uptake variance using the proposed method, we computed the CRLB for the estimates with different ranges of energy windows. 
In this experiment, we considered four energy window ranges: single window 1, windows 1-2, windows 1-3, and windows 1-4, as depicted in Fig.~\ref{fig:emi_spect}.
The CRLB with more energy window ranges are considered in the Supplementary material and Li et al.~\cite{liGitHub}.
To compute the CRLB, we first generated a patient phantom with the average patient size, a 33.75 mm diameter lesion, standard total uptake, and regional uptake ratios as shown in Table~\ref{tb:uptake}.
Calculating the CRLB for the estimated uptake with different numbers of energy windows requires the corresponding system matrix of the patient. These system matrices were generated for both isotopes, as described in Sec.~\ref{sec:implementation}.
The corresponding CRLB-derived NSD values were then calculated.

As shown in the results presented later (Fig.~\ref{fig:CRLB_win}), the CRLB was the lowest when data from all four energy windows were used. To evaluate if the MEW-PDQ method could yield estimates with variances that approach this lowest theoretical limit for all VOIs, we generated 50 noise realizations of the patient phantom described above in this section. With each noise realization, the MEW-PDQ method was applied to estimate the mean regional uptake of both $\mathrm{^{223}Ra}$ and $\mathrm{^{227}Th}$ from projections in all four energy windows (Fig.~\ref{fig:emi_spect}).
The NSDs of these estimates were compared to the CRLB-derived NSD values with all these four energy windows.

\subsubsection{Evaluation using a virtual imaging trial (VIT)}
\label{sec:VIT}
VITs provide a rigorous and objective mechanism to evaluate the performance of new imaging technologies in simulated clinical scenarios that model patient population variability~\cite{abadi2020virtual}. In our VIT, we generated simulations for 60 realistic patients, taking into account variations in body size, lesion diameter, lesion locations, and regional uptake ratios. The heights and weights of the patients were sampled from demographic data of 2718 US male adults obtained from the 2015-16 National Health and Nutrition Examination Survey (NHANES) of the Centers of Disease Control and Prevention (CDC). Next, based on clinically derived parameters~\cite{murray2017potential}, the lesion diameter was sampled from a Gaussian distribution with a mean of 33.75 mm and a standard deviation of 12.64 mm. Lesion locations were randomly assigned within the pelvic bones of the patients, a major site for bone metastases in prostate cancer patients. We simulated a scenario where the patients were initially administered with 11~MBq activity of $\mathrm{^{227}Th}$ and imaged 82~h after administration, when the $\mathrm{^{227}Th}$ to $\mathrm{^{223}Ra}$ uptake ratio in the patient was on average approximately 5:1. The activity uptake of both isotopes in the four VOIs were sampled independently from a normal distribution, with the mean regional uptake ratios shown in Table~\ref{tb:uptake} and a 10\% standard deviation. We generated a single noise realization for each simulated patient.

\subsubsection{Sensitivity to intra-regional uptake heterogeneities}
As we note from Eq.~\eqref{eq:voi_def}, the MEW-PDQ method is based on an object model where the different VOIs have constant uptake. Thus, the method does not explicitly model potential intra-regional uptake heterogeneities. However, in clinical scenarios, intra-regional uptake heterogeneity may be present. One approach to mitigate this limitation is to use both CT and reconstructed SPECT images to deliberately define VOIs with relatively homogeneous uptake. However, even with this approach, due to the limited resolution and high noise levels of SPECT, minor intra-regional uptake heterogeneities may be obscured on the images. Further, intra-regional heterogeneity could be present at a continuous level, and thus, some amount of heterogeneity may be present within the different VOIs. Hence, we assessed the performance of the methods in the presence of these heterogeneities.

In this experiment, patients with heterogeneous uptake distributions within three large regions: the background, gut, and lesion, were simulated. To simulate intra-regional uptake heterogeneity at a continuous level, we modeled the activity uptake in these regions using a 3D lumpy model~\cite{liu2021observer}.
Specifically, for region $k$ with support $\phi_{k}^{VOI}\left( \mathbf{r} \right)$, as defined in Eq.~\eqref{eq:dif_roi}, the intra-regional uptake distribution, denoted by $f_{k}\left( \mathbf{r} \right)$, was formulated as:
\begin{equation}
    f_{k}(\bm{r}) = {a_{k}\phi}_{k}^{VOI}\left( \bm{r} \right)\sum_{p = 1}^{P_{k}}\frac{1}{\left( \sigma_{k}\sqrt{2\pi} \right)^{3}}\exp\left\lbrack - \frac{\left( \bm{r} - \bm{c}_{\bm{p}} \right)^{2}}{2\sigma_{k}^{2}} \right\rbrack,
\end{equation}
where $P_{k}$, $a_{k}$, and $\sigma_{k}$ denote the total number, magnitude, and width of lumps in region $k$, respectively, and $\mathbf{c}_{\mathbf{p}}$ represents the center of the $p^{th}$ lump. Each $\mathbf{c}_{\mathbf{p}}$ was randomly sampled with uniform probability for each spatial location within a given region. The value of $a_{k}$ was chosen such that the average uptake activity in region $k$ matched the designed mean uptake value, $\lambda_{k}$. Finally, $P_{k}$ is sampled from a Poisson distribution with a mean of $\overline{P_{k}}$.

\begin{figure}
    \centering
    \includegraphics[width=0.48\textwidth]{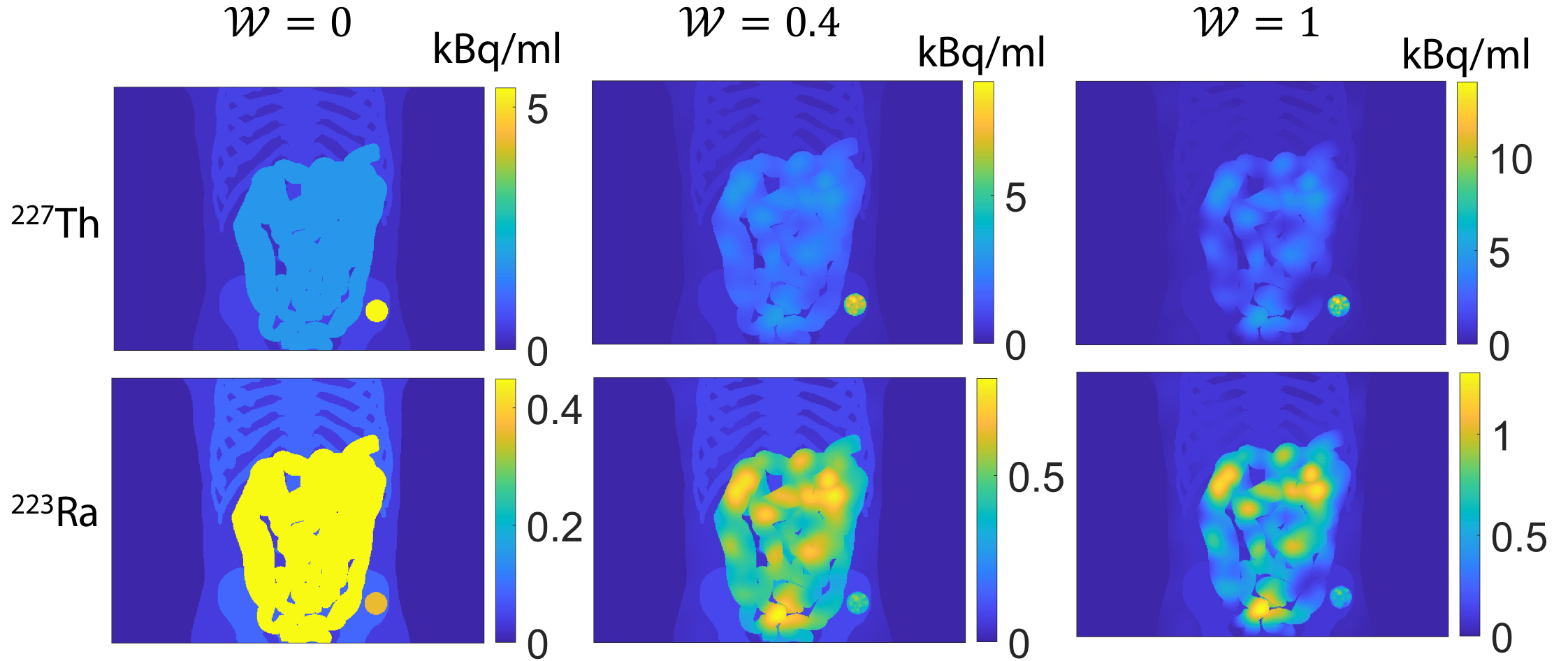}
    \caption{The maximum intensity projections of the activity maps for a representative patient with homogeneous intra-regional uptake ($\mathcal{W} = 0$), moderate intra-regional heterogeneity ($\mathcal{W} = 0.4$), and a fully heterogeneous patient ($\mathcal{W} = 1$).}
    \label{fig:Het_pat_level}
\end{figure}

Our goal was to study the performance of the proposed method across a range of intra-regional heterogeneities. We first simulated patients with high intra-regional uptake heterogeneity. This was done by considering a large mean number of lumps (i.e. large $\overline{P_{k}}$) and with a small lump width (i.e. small $\sigma_{k}$) for each region. Specifically, for the background, gut, and lesion regions, the mean number of lumps, $\overline{P_{k}}$, was set to 380, 140, and 490 respectively, while the lump width, $\sigma_{k}$, was modeled as 23 mm, 16 mm, and 2 mm, respectively. We refer to these generated patients as fully heterogeneous patients. Denote the activity uptake distribution in one of the fully heterogeneous patients as $f^{het}\left( \mathbf{r} \right).$ Next, we simulated the same patient with the same mean regional uptake but homogeneous uptake distribution in the different regions. Denote this uptake distribution by $f^{\hom}\left( \mathbf{r} \right)$. We used these distributions to generate varying levels of heterogeneity as per the following mixture model:
\begin{equation}
    \label{eq:het_level_pat_gen}
    f_{\mathcal{W}}^{Mix}\left( \bm{r} \right) = \mathcal{W}f^{het}\left( \bm{r} \right) + \left( 1 - \mathcal{W} \right)f^{\hom}\left( \bm{r} \right).
\end{equation}
where the values of $\mathcal{W}$ were chosen as 0.1, 0.2, 0.3, 0.4, and 0.8, respectively. A higher $\mathcal{W}$ value led to higher intra-regional heterogeneity. Fig.~\ref{fig:Het_pat_level} shows the maximum intensity projection of the activity map of a representative patient with homogeneous intra-regional uptake ($\mathcal{W} = 0$), moderate intra-regional heterogeneity ($\mathcal{W} = 0.4$), and the fully heterogeneous patient ($\mathcal{W} = 1$).

To investigate the performance of the method across a range of patients, we simulated five patients at each considered level of intra-regional heterogeneity, where all patients had an average body size, an identically positioned pelvic bone lesion with a 33.75 mm diameter, standard total uptake, and mean regional uptake distributions as present in Table~\ref{tb:uptake}. For each simulated patient, we generated 50 noise realizations.

\subsubsection{Sensitivity to inaccuracy in VOI definitions}
Both conventional RBQ methods and the proposed method rely on VOI definitions for estimating regional uptake. Typically, these VOIs are derived from patient CT scans, acquired sequentially with SPECT scans. While aligning CT and SPECT data is feasible, inaccuracies in VOI definitions can still arise. In this study, we evaluated the performance of the methods in the presence of such inaccuracies. We considered inaccuracies in VOIs originating from two sources: rigid-body transformations, such as patient movement between CT and SPECT scans, persisting even after registration~\cite{tang2006co,song2010effect}, and non-rigid transformations occurring during SPECT scanning, like patient breathing.

In experiments described in this section, we considered a patient with average body size and a pelvic bone lesion of 33.75 mm diameter. The phantom had standard total uptake and mean regional uptake distributions as specified in Table~\ref{tb:uptake}.

\begin{figure}
    \centering
    \includegraphics[width=0.35\textwidth]{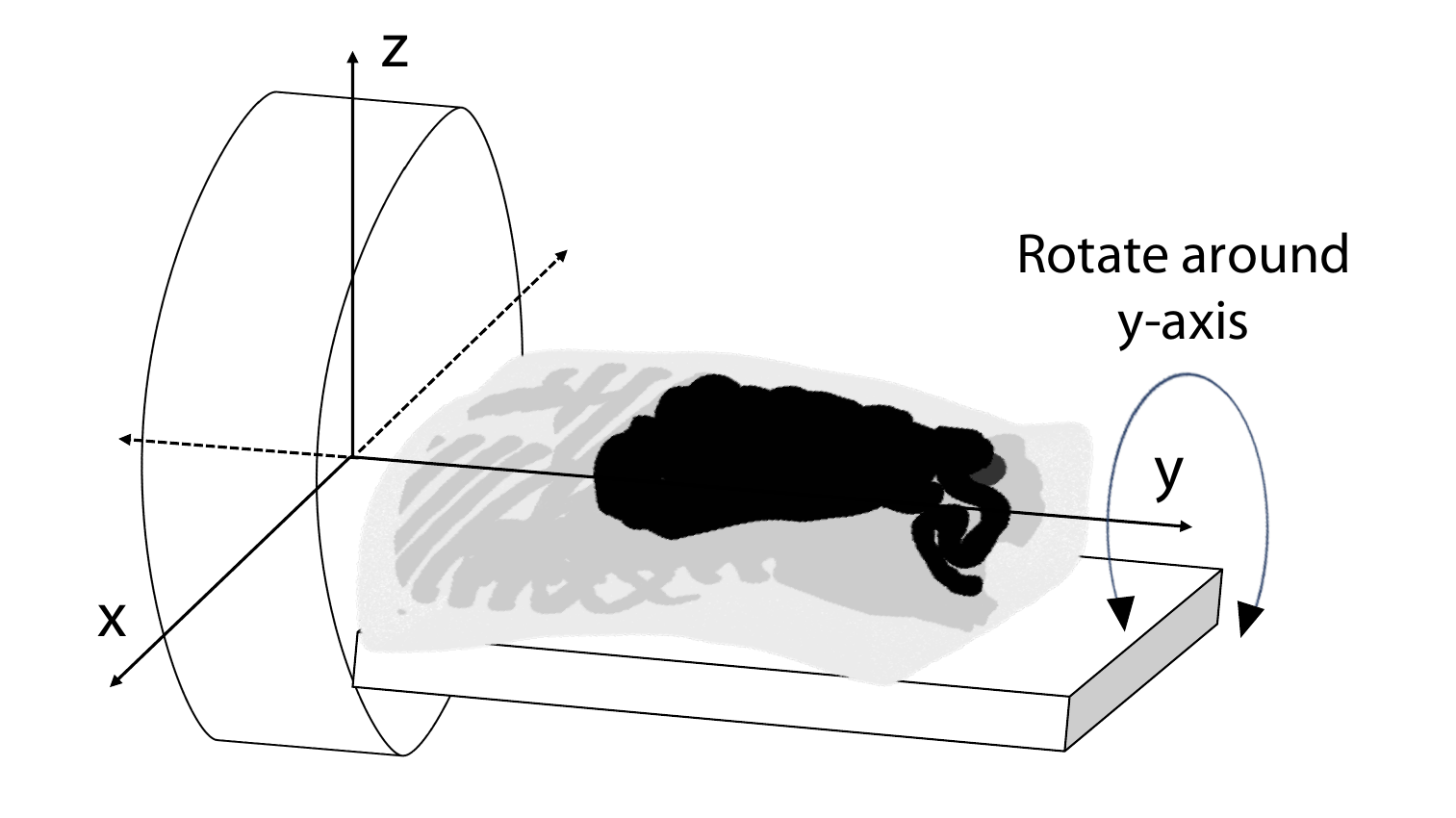}
    \caption{Illustration of the coordinates used in simulating the rigid misregistration in VOI definitions.}
    \label{fig:mismatch_coordinate}
\end{figure}

First, we evaluate the performance of the methods with inaccuracies in VOI definitions due to rigid-body transformations. We assumed that the CT scan of the patient was conducted upfront, yielding the VOIs. Then, we simulated shifting and rotation of the patient as illustrated in Fig.~\ref{fig:mismatch_coordinate} to model misregistration between the SPECT and CT scans. In previous studies, it has been observed that even after the application of computer-aided techniques to align the reconstructed SPECT and CT images, certain levels of misalignment remain~\cite{tang2006co,song2010effect}.
We simulated those residual misalignments between the SPECT and CT scans. Specifically, the patient was shifted by 2, 4, 6, and 9 voxels in both directions along the x-axis, and by 2, 4, and 6 voxels in both directions along the y-axis, respectively. Each voxel had a side length of 1.105 mm. The patient was also rotated by 1 and 2 degrees in both directions around the y-axis. SPECT projections corresponding to the shifted and rotated patient were generated. We generated 50 noise realizations for each simulated misregistration. Both the MEW-PDQ and conventional RBQ methods used the attenuation map and the VOI definitions as obtained from the CT scan to process the projection data and estimate the regional uptake.

Next, we evaluate the performance of the methods with inaccuracies in VOI definitions due to non-rigid transformations. We considered scenarios where patients held their breath during the CT scan in a fully inhaled position, and breathed normally with a five-second respiration cycle during the SPECT scan. The XCAT software enables simulation of this breathing motion. For each respiration cycle, we sampled the XCAT-generated activity and attenuation maps at one-second time intervals, resulting in five temporal frames per respiratory cycle. The SPECT projections for each of these five frames were then added to generate the average projection data for one respiratory cycle. This study design is similar to a previous study assessing the impact of respiratory motion on detecting defects in myocardial perfusion SPECT~\cite{yang2009evaluation}.
50 noise realizations of the final SPECT projection were generated. Subsequently, we applied the MEW-PDQ and RBQ methods, using the VOI definitions and attenuation maps as obtained from the CT scan.

\section{Results}
\subsection{Convergence of the proposed method}
\begin{figure}
    \centering
    \includegraphics[width=0.45\textwidth]{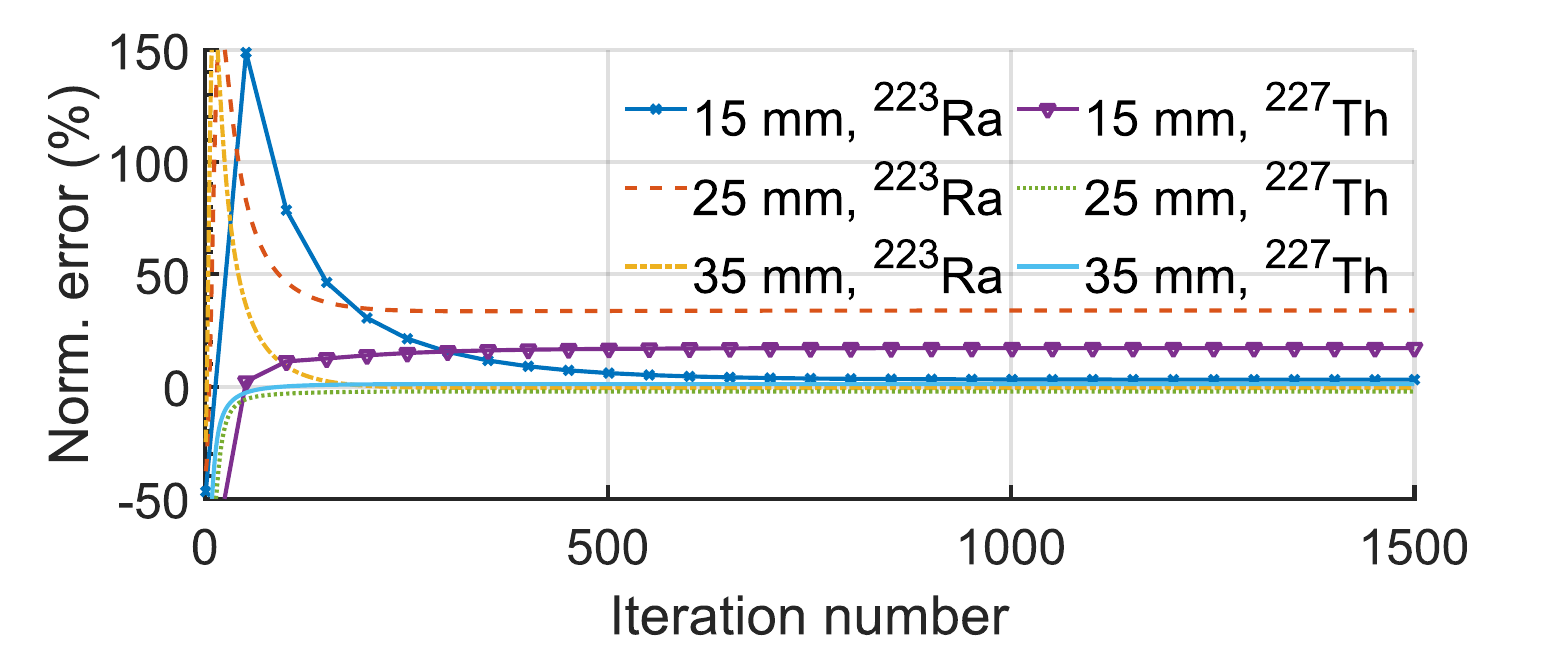}
    \caption{Normalized error in the estimated lesion uptake of $\mathrm{^{227}Th}$ and $\mathrm{^{223}Ra}$ with different lesion diameters as a function of the number of iterations, showing the convergence of the MEW-PDQ method.}
    \label{fig:convergence}
\end{figure}

Fig.~\ref{fig:convergence} shows the normalized error in the estimated lesion uptake of $\mathrm{^{227}Th}$ and $\mathrm{^{223}Ra}$ as a function of the number of iterations of the MEW-PDQ method. The method converged after 1000 iterations for both isotopes and all three lesion diameters. Thus, we chose 1000 as the number of iterations for subsequent experiments using the MEW-PDQ method. 

\subsection{Quantitation performance for different lesion sizes and contrasts}
\begin{figure}
    \centering
    \includegraphics[width=0.48\textwidth]{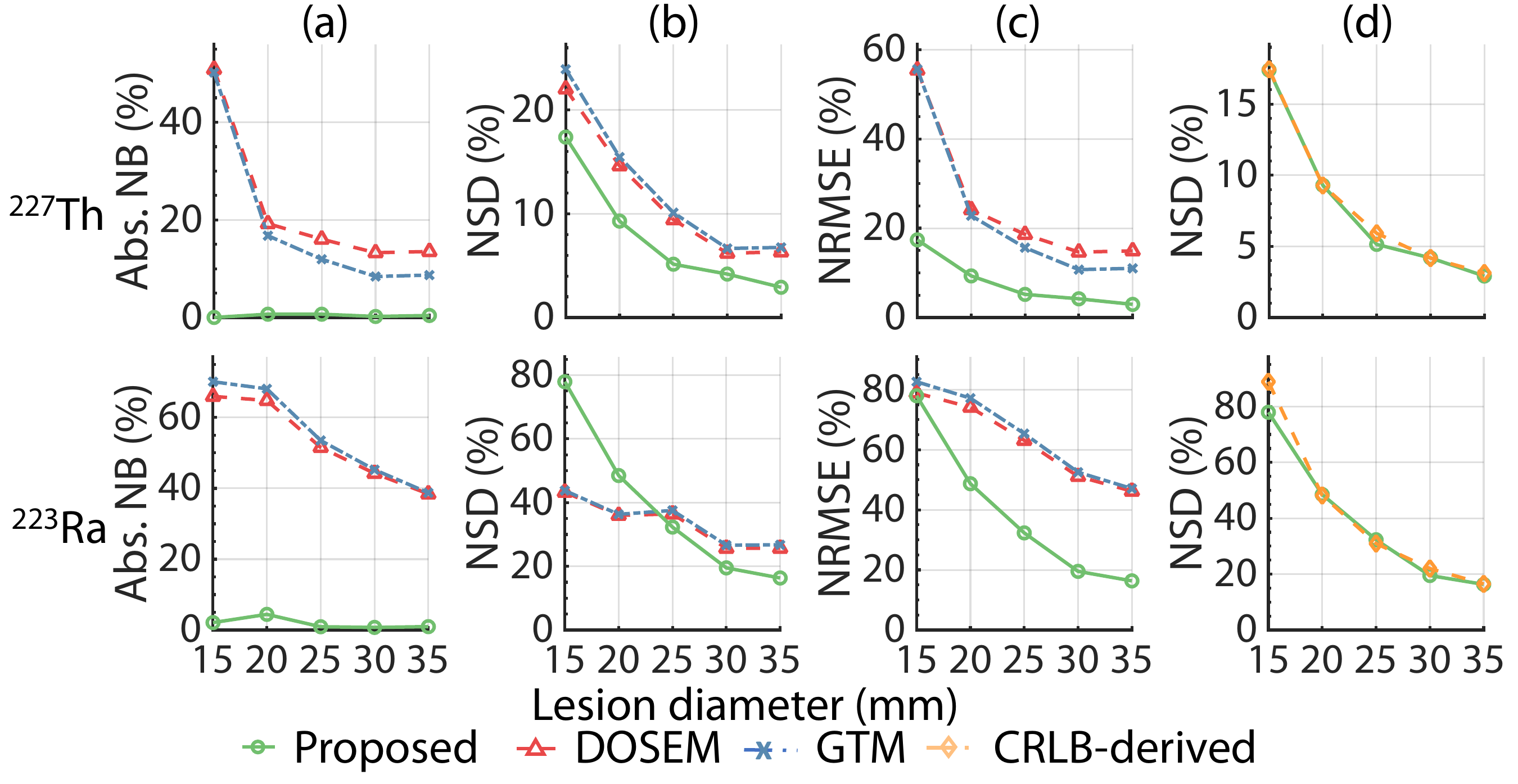}
    \caption{The (a) absolute NB, (b) NSD, and (c) NRMSE of the estimated lesion uptake of $\mathrm{^{227}Th}$ and $\mathrm{^{223}Ra}$ as a function of lesion diameter; (d) the comparisons of the NSD yielded by the proposed method with that derived from the CRLB.}
    \label{fig:size}
\end{figure}

\begin{figure}
    \centering
    \includegraphics[width=0.48\textwidth]{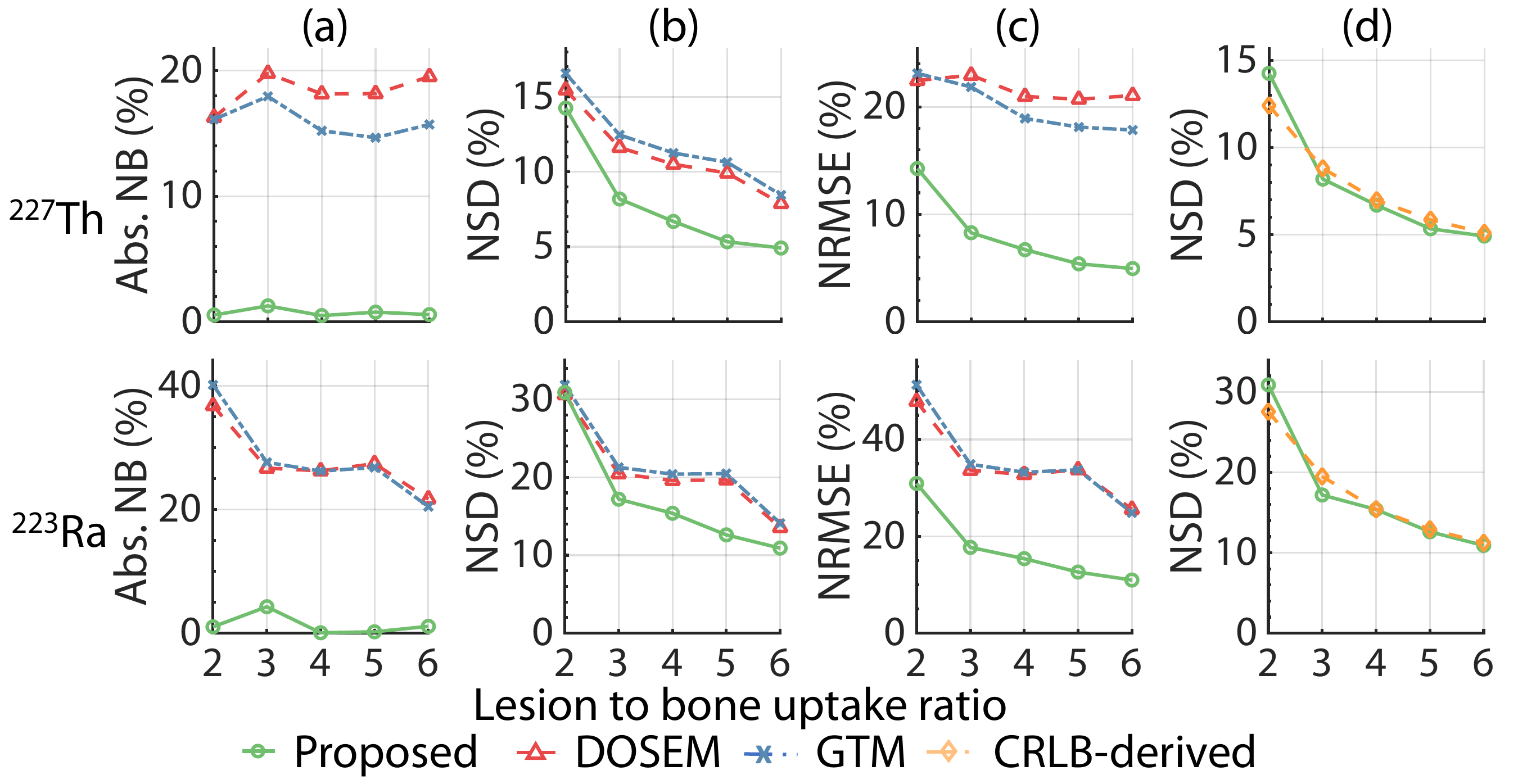}
    \caption{The (a) absolute NB, (b) NSD, and (c) NRMSE of the estimated lesion uptake of $\mathrm{^{227}Th}$ and $\mathrm{^{223}Ra}$ as a function of lesion-to-bone uptake ratio; (d) the comparisons of the NSD yielded by the proposed method with that derived from the CRLB.}
    \label{fig:ratio}
\end{figure}

The absolute NB, NSD, and NRMSE of the estimated lesion uptake as a function of lesion diameter and LBUR are shown in Figs.~\ref{fig:size} and~\ref{fig:ratio}, respectively. We also compared the NSD of the estimates yielded by the proposed method with CRLB-derived NSD values. For both $\mathrm{^{227}Th}$ and $\mathrm{^{223}Ra}$ in lesions with all diameters and LBURs, the MEW-PDQ method consistently yielded close to zero bias and an NSD close to that derived from the CRLB. The proposed method also outperformed the DOSEM and GTM-based methods for all the considered lesion diameters and the LBURs.

\subsection{Quantitation performance for different~$\mathrm{^{227}Th}$~to~$\mathrm{^{223}Ra}$ uptake ratios}
\begin{figure}
    \centering
    \includegraphics[width=0.48\textwidth]{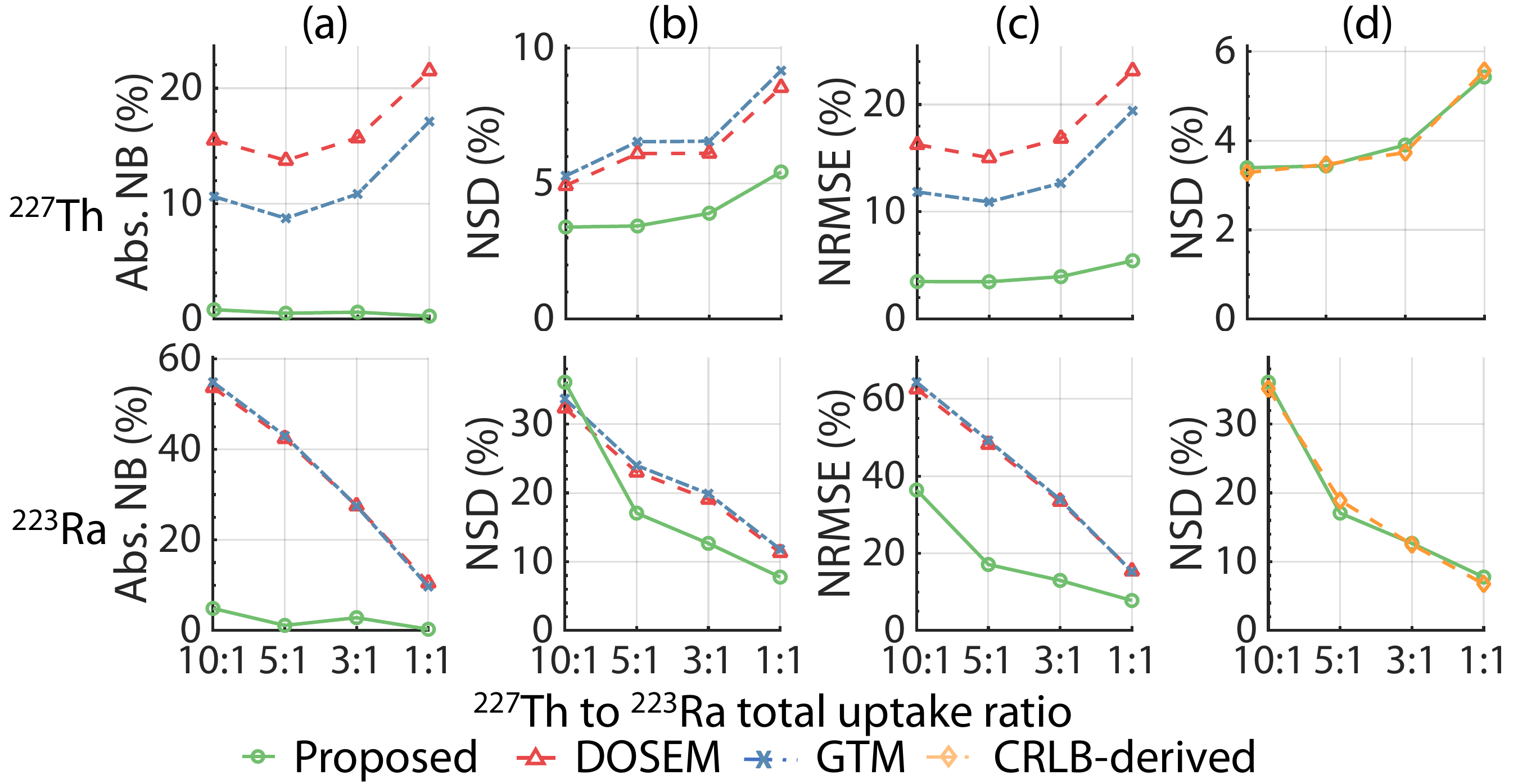}
    \caption{The (a) absolute NB (b) NSD, and (c) NRMSE of the estimated lesion uptake of $\mathrm{^{227}Th}$ and $\mathrm{^{223}Ra}$ as a function of $\mathrm{^{227}Th}$~to~$\mathrm{^{223}Ra}$ uptake ratio; (d) the comparisons of the NSD yielded by the proposed method with that derived from the CRLB.}
    \label{fig:timeAfter}
\end{figure}

Fig.~\ref{fig:timeAfter} shows the absolute NB, NSD, and NRMSE of the estimated lesion uptake for different $\mathrm{^{227}Th}$~to~$\mathrm{^{223}Ra}$ uptake ratios.
For all $\mathrm{^{227}Th}$~to~$\mathrm{^{223}Ra}$ uptake ratios,
the MEW-PDQ method consistently yielded almost zero NB and an NSD close to the CRLB-derived NSD value. Further, the method outperformed both the DOSEM and GTM methods in terms of accuracy and overall error for all considered $\mathrm{^{227}Th}$~to~$\mathrm{^{223}Ra}$ uptake ratios.

\subsection{Impact of using multiple energy windows on the precision of estimates}
\label{sec:impact_usingMEW}
\begin{figure}
    \centering
    \includegraphics[width=0.45\textwidth]{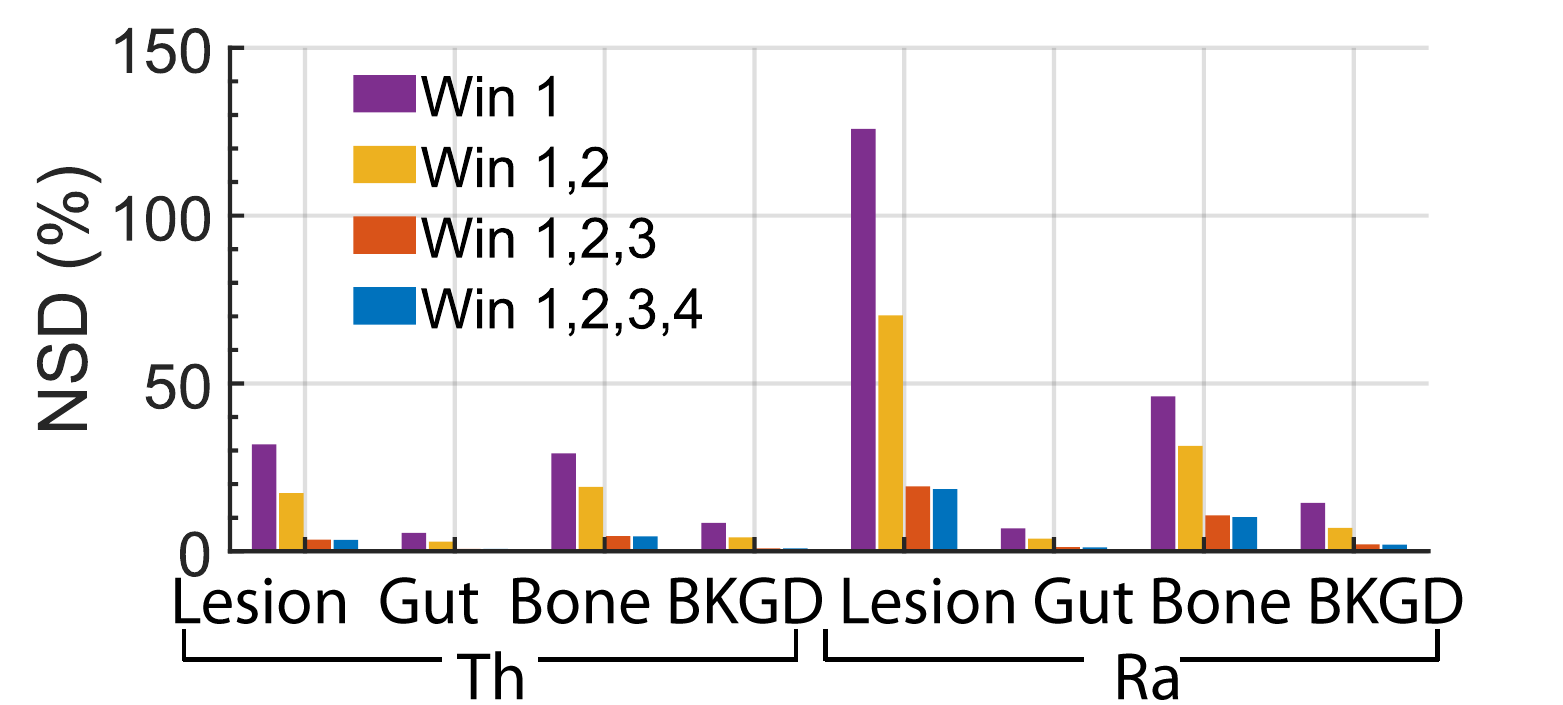}
    \caption{Comparing the CRLB-derived NSDs of the estimated regional uptake of $\mathrm{^{227}Th}$ and $\mathrm{^{223}Ra}$, for different numbers of energy windows.}
    \label{fig:CRLB_win}
\end{figure}

Fig.~\ref{fig:CRLB_win} shows the CRLB-derived NSDs using projection data from different numbers of energy windows. The CRLB-derived NSD decreased with an increase in the number of energy windows. Compared with using only one energy window, using all four windows led to at least a 78\% decrease in the CRLB-derived NSD values for all regions and isotopes. 

\begin{figure}
    \centering
    \includegraphics[width=0.45\textwidth]{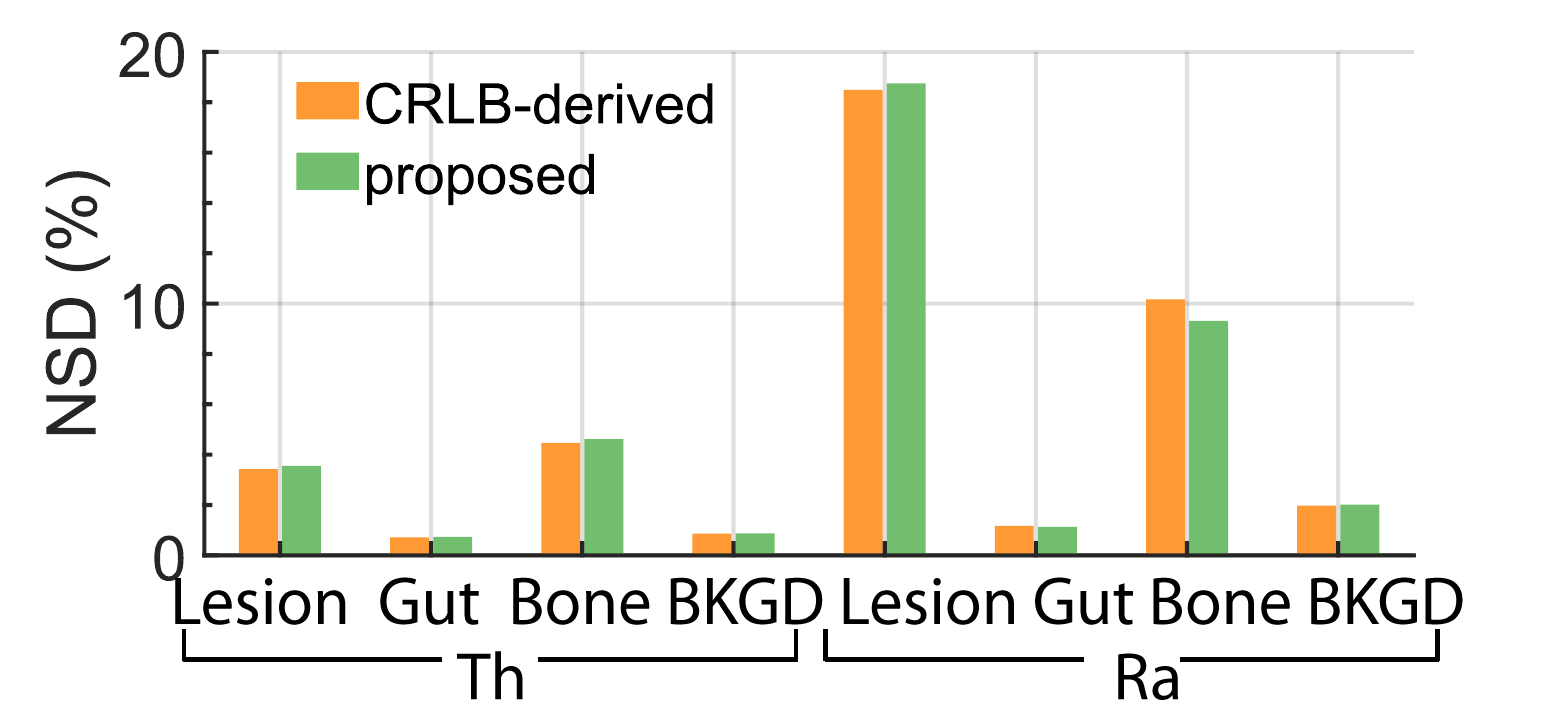}
    \caption{NSD of the estimated regional uptake of $\mathrm{^{227}Th}$ and $\mathrm{^{223}Ra}$ using the MEW-PDQ method compared with the CRLB-derived NSDs.}
    \label{fig:CRLB}
\end{figure}

Fig.~\ref{fig:CRLB} compares the NSD of the estimated regional uptake of $\mathrm{^{227}Th}$ and $\mathrm{^{223}Ra}$ using the MEW-PDQ method with the CRLB-derived NSDs. For both cases, projection data from all four energy windows were used. We observe that the proposed method yielded an NSD very close to that derived from the CRLB for all the considered regions and both isotopes.

\subsection{Virtual imaging trial} 
\begin{figure}
    \centering
    \includegraphics[width=0.45\textwidth]{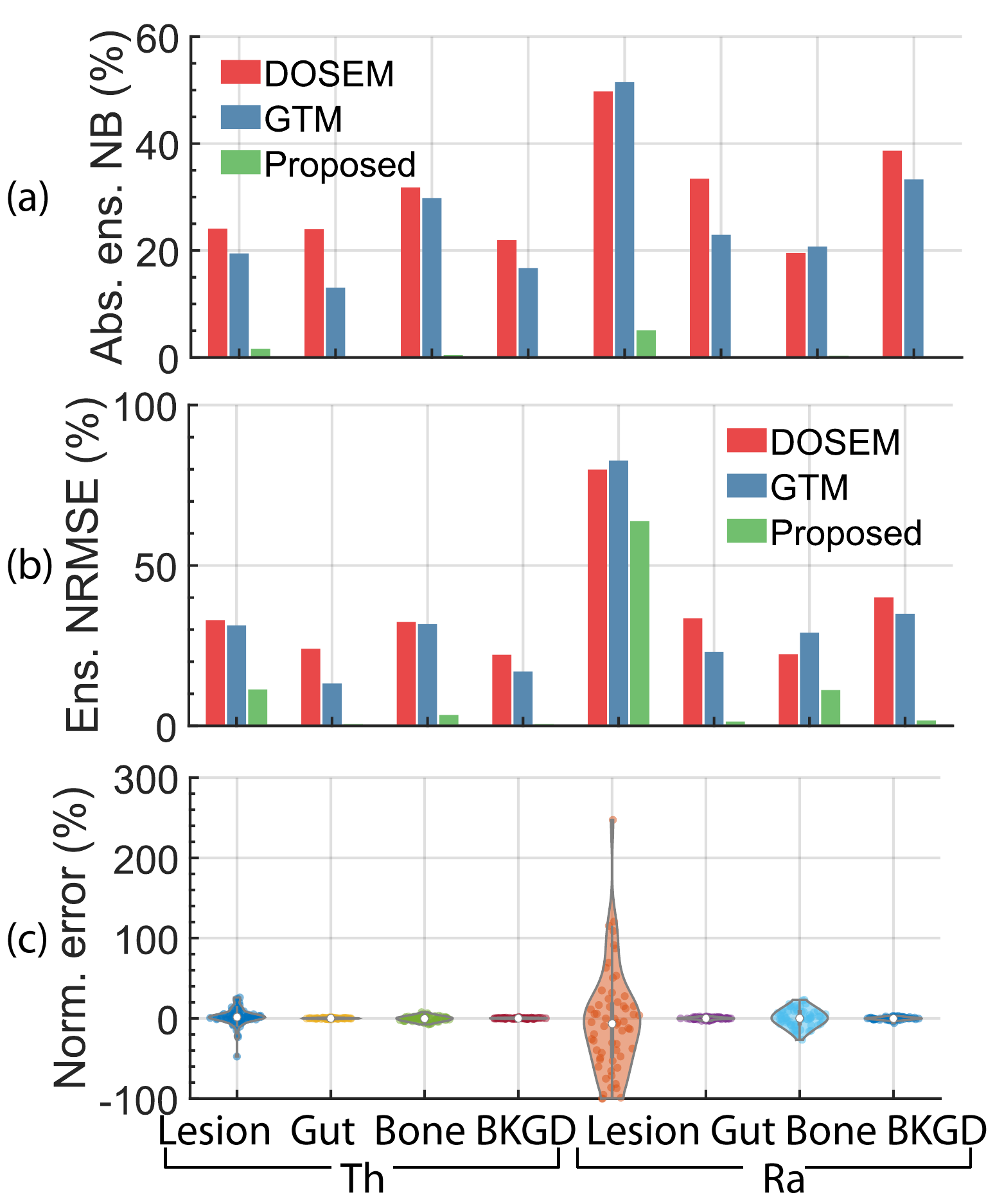}
    \caption{The (a) absolute ensemble NB and (b) ensemble NRMSE of the estimated uptake of $\mathrm{^{227}Th}$ and $\mathrm{^{223}Ra}$ across different regions using the different methods in the VIT. (c) A violin plot showing the distribution of the normalized error across sixty patients.}
    \label{fig:VIT}
\end{figure}

Fig.~\ref{fig:VIT} a and b show the absolute ensemble NB and ensemble NRMSE of the estimated regional uptake of $\mathrm{^{227}Th}$ and $\mathrm{^{223}Ra}$ in the lesion, gut, bone, and background in the VIT. We observe that the method consistently outperformed the DOSEM and GTM-based methods on the task of estimating the uptake of both isotopes in all regions, considering both accuracy and overall error, and in fact, yielded at least 10 times lower ensemble bias in estimating lesion uptake than these methods. Further, the absolute ensemble NB obtained with the MEW-PDQ method for both isotopes in all regions was consistently lower than 5.5\%. The accurate performance was observed for individual patients, as we observe in Fig.~\ref{fig:VIT} c, which shows the distribution of the normalized error between the true and estimated uptake for all 60 patients using the proposed MEW-PDQ method.

\subsection{Sensitivity to intra-regional uptake heterogeneities}
\begin{figure}
    \centering
    \includegraphics[width=0.48\textwidth]{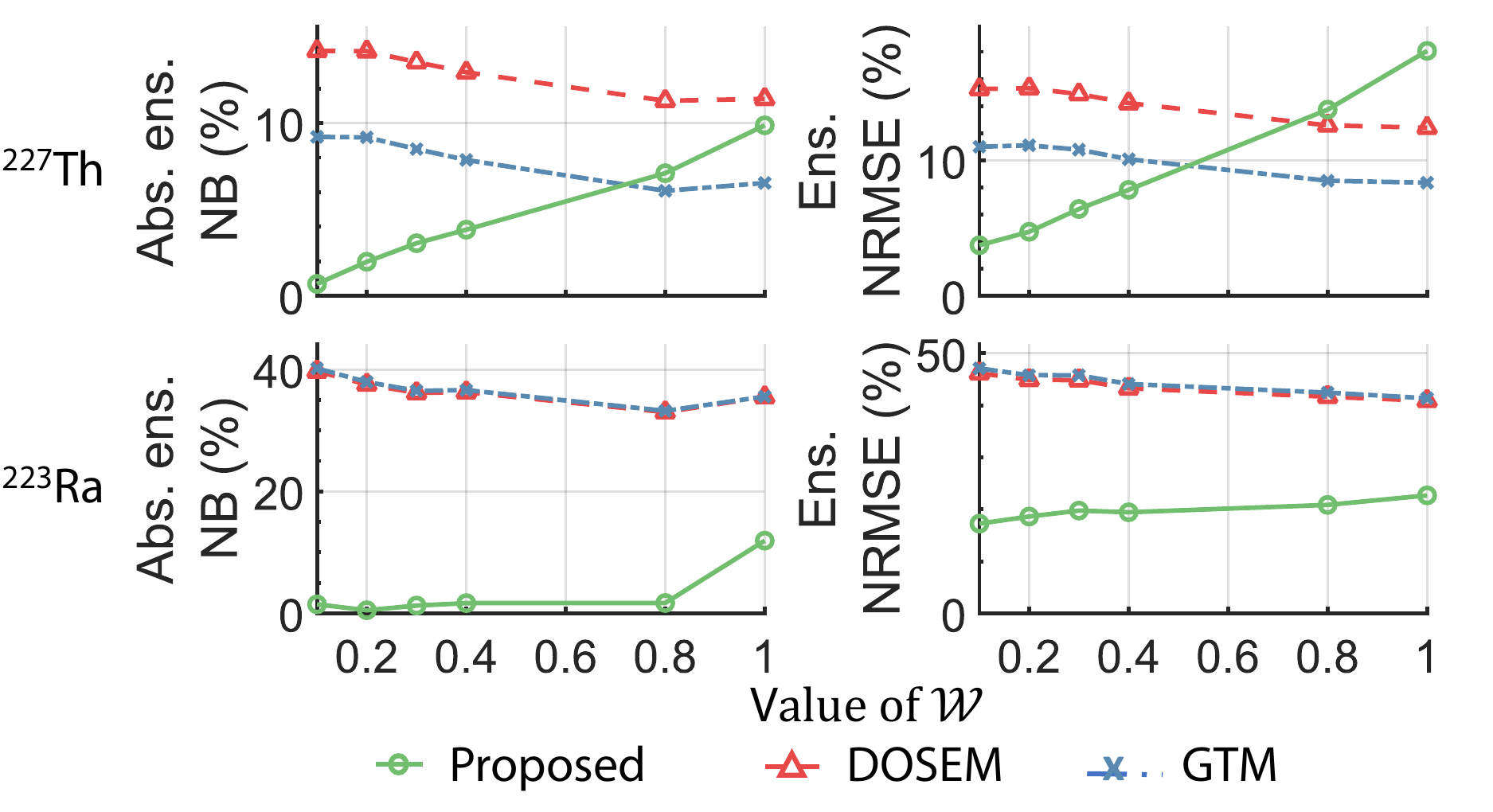} 
    \caption{The absolute ensemble NB and ensemble NRMSE of the estimated lesion uptake of $\mathrm{^{227}Th}$ and $\mathrm{^{223}Ra}$ across patients generated by setting $\mathcal{W}$ to each of the values in Eq.~\eqref{eq:het_level_pat_gen}. A higher $\mathcal{W}$ value indicates that the patients have a higher level of intra-regional heterogeneity.}
    \label{fig:het_box_plot}
\end{figure}

Fig.~\ref{fig:het_box_plot} shows the absolute ensemble NB and ensemble NRMSE of the estimated lesion uptake of $\mathrm{^{227}Th}$ and $\mathrm{^{223}Ra}$ across patients for different levels of intra-regional heterogeneity, as generated by varying $\mathcal{W}$ in Eq.~\eqref{eq:het_level_pat_gen}.
We found that when $\mathcal{W} \leq 0.4$, the proposed method outperformed the conventional RBQ methods in both accuracy and overall error. Further, even for higher values of $\mathcal{W}$, the proposed method outperformed the conventional RBQ methods in quantifying the lesion uptake of $\mathrm{^{223}Ra}$.

\subsection{Sensitivity to inaccuracy in VOI definitions}
\begin{figure}
    \centering
    \includegraphics[width=0.48\textwidth]{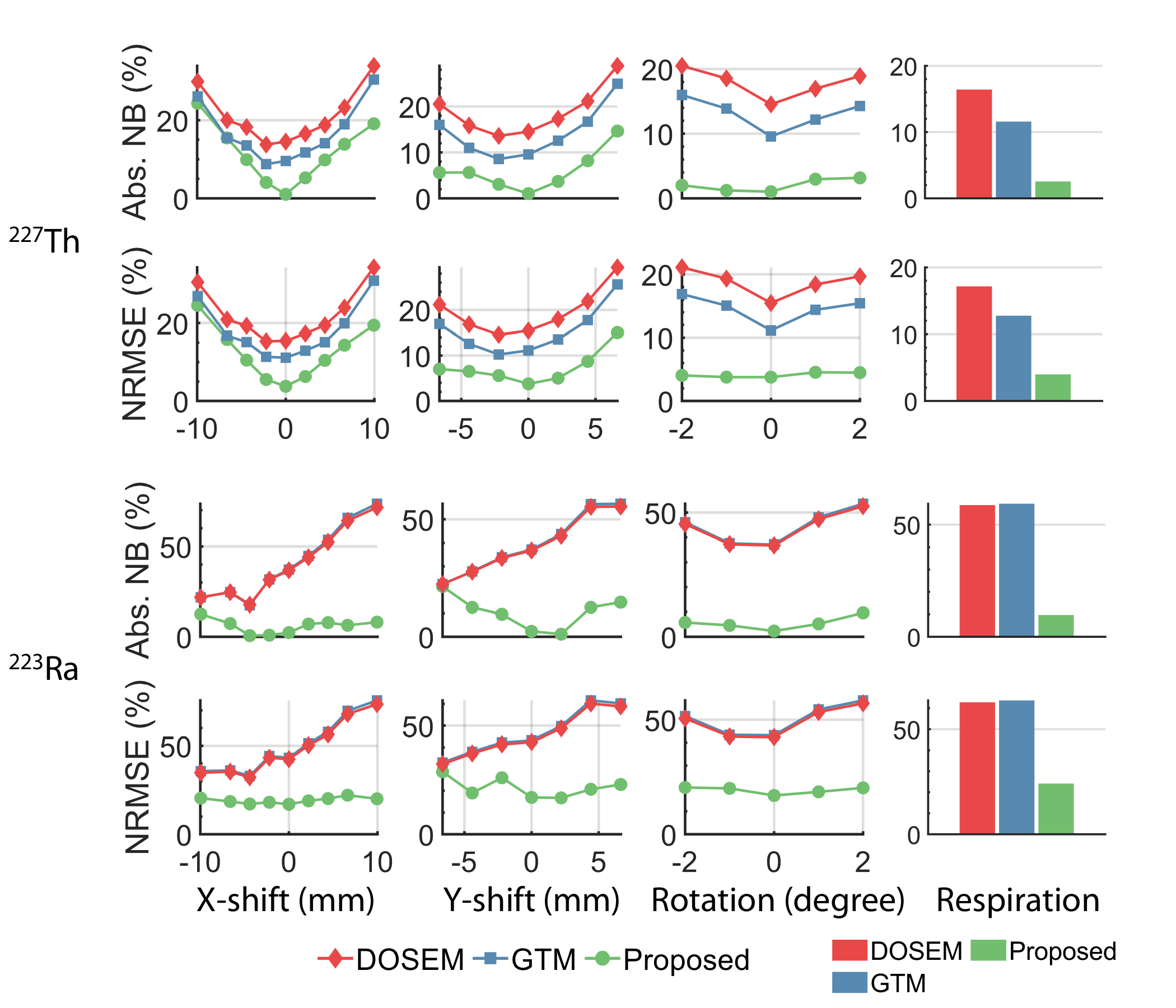}
    \caption{The absolute NB and NRMSE of estimated lesion uptake of $\mathrm{^{227}Th}$ and $\mathrm{^{223}Ra}$ with the presence of each considered inaccuracy in VOI definitions.}
    \label{fig:mismatch}
\end{figure}

Fig.~\ref{fig:mismatch} presents the absolute NB and NRMSE of the estimated lesion uptake of both $\mathrm{^{227}Th}$ and $\mathrm{^{223}Ra}$ for varying levels and types of inaccuracy in VOI definitions. Across all these inaccuracies, the proposed MEW-PDQ method consistently demonstrated better performance over the conventional RBQ methods in terms of both accuracy and overall error.

\section{Discussion}
\label{sec:dis}
We have designed, implemented, and evaluated a MEW-PDQ method to jointly quantify the activity uptake of $\mathrm{^{227}Th}$ and $\mathrm{^{223}Ra}$ in $\mathrm{^{227}Th}$-based $\alpha$-RPTs. The method was designed to address the challenges in quantitative SPECT for $\mathrm{^{227}Th}$-based $\alpha$-RPTs, including the extremely low counts, crosstalk contaminations between $\mathrm{^{227}Th}$ and $\mathrm{^{223}Ra}$, and the complicated isotope and SPECT physics. The results (Figs.~\ref{fig:size}-\ref{fig:mismatch}) demonstrate the efficacy of the method in jointly estimating the regional uptake of $\mathrm{^{227}Th}$ and $\mathrm{^{223}Ra}$ in multiple scenarios.

The results in Figs.~\ref{fig:size}-\ref{fig:timeAfter} demonstrate that the MEW-PDQ method yielded accurate estimates of lesion uptake of both $\mathrm{^{227}Th}$ and $\mathrm{^{223}Ra}$ and outperformed the DOSEM and GTM-based methods with varying lesion characteristics, including different lesion sizes (Fig.~\ref{fig:size} a), LBURs (Fig.~\ref{fig:ratio} a), and $\mathrm{^{227}Th}$ to $\mathrm{^{223}Ra}$ uptake ratios (Fig.~\ref{fig:timeAfter} a) in the case where there was homogeneous uptake in the VOIs, consistent with the assumption of the proposed method. 
The close-to-zero bias estimates for both isotopes highlight the efficacy of the proposed method in accounting for the crosstalk contamination, one of the major image-degrading effects in quantitative SPECT for $\mathrm{^{227}Th}$-based RPTs. Further, the proposed method was observed to be relatively insensitive to PVEs, as demonstrated by the high accuracy for even small lesion sizes. We note that the proposed method performs direct ML estimation of the targeted VOI uptake, and ML estimation is known to be asymptotically unbiased. In addition, the proposed method yielded NSD values close to those derived from CRLBs (Figs.~\ref{fig:size}-\ref{fig:timeAfter} b and \ref{fig:CRLB}). Again, the proposed method is an ML estimator. If an efficient estimator exists, the ML estimator is efficient~\cite{barrett2013foundations}. Thus, these results indicate that the MEW-PDQ method may be an efficient estimator.

An important feature of the MEW-PDQ method is the use of multiple energy windows in quantification. As theoretically shown through the application of the data-processing inequality, discarding projections in one or more energy windows decreases the Fisher information for estimating parameters from the acquired data~\cite{chun2019algorithms}.
This theoretical finding is manifested in our results in Fig.~\ref{fig:CRLB_win}, where we see that using data from multiple energy windows reduces the CRLB significantly. This further motivates the need to develop quantitative SPECT methods that can use data from multiple energy windows, such as the proposed method. This is particularly important for $\alpha$-particle SPECT, as the low number of detected counts can be increased by including data from multiple energy windows. Further, as shown in Fig.~\ref{fig:CRLB}, using projections in all four energy windows, the variance of estimates obtained by the proposed method approached the corresponding CRLB for all VOIs and both isotopes. This finding demonstrates the effectiveness of the proposed method in extracting information from data contained in these energy windows. Overall, these results show that effectively using data from multiple energy windows in jointly quantifying regional uptake of $\mathrm{^{227}Th}$ and $\mathrm{^{223}Ra}$, as can be enabled by the proposed method, can yield improved quantification.

The evaluation of the method in the VIT (Fig.~\ref{fig:VIT}) demonstrated the efficacy of the method across a population of patients. We see that the MEW-PDQ method yielded averaged absolute ensemble NB and ensemble NRMSE values as low as 1.0\% and 11.7\%, respectively (Fig.~\ref{fig:VIT}). In contrast, the DOSEM-based method yielded corresponding average absolute ensemble NB and ensemble NRMSE of 30.4\% and 35.9\%, respectively. The GTM-based method yielded corresponding values of 25.9\% and 32.9\%, respectively. These observations for the DOSEM and GTM-based method are consistent with previous literature~\cite{ghaly2019quantitative,li2022projection,benabdallah2019223,yue2016f,gustafsson2020feasibility} and confirm the limitation of these conventional approaches for the task of joint $\mathrm{^{227}Th}$ and $\mathrm{^{223}Ra}$ regional uptake quantification. These results further demonstrate the reliability of the proposed MEW-PDQ method for this task.

The proposed MEW-PDQ method does not account for spatial intra-regional uptake heterogeneity, leading to model mismatches in scenarios where such heterogeneity is present. As illustrated in Fig.~\ref{fig:het_box_plot}, as the level of intra-regional uptake heterogeneity increases, the absolute NB and NRMSE of the estimated lesion uptake from the proposed method also increase. Despite this, with moderate levels of intra-regional heterogeneities, the proposed method still consistently outperforms the conventional RBQ methods. This finding suggests that the MEW-PDQ method could still be a reliable choice with better performance than conventional RBQ methods given such moderate levels of intra-regional heterogeneity. However, given high levels of intra-regional heterogeneity, the proposed method may yield less reliable estimates of regional uptake compared to the conventional RBQ methods (Fig.~\ref{fig:het_box_plot}). Under those conditions, the model mismatch for the proposed method in representing the activity uptake distribution within VOIs may impact the performance of the method. We note that the actual extent of intra-regional heterogeneity in patients undergoing $\mathrm{^{227}Th}$-based $\alpha$-RPTs remains unclear. The range of intra-regional uptake heterogeneities evaluated in this study was deliberately extensive to ensure a comprehensive understanding of the methods in various conditions, including those that pose significant challenges. Additionally, to leverage the reliable performance of the proposed method in clinical applications, a practical strategy is to carefully delineate VOIs using both CT and reconstructed SPECT images to define VOIs with low intra-regional heterogeneities. However, not accounting for intra-regional heterogeneity remains a limitation of the proposed method when defining VOIs with relatively homogeneous uptake is not achievable. To address this limitation, a Wiener estimator-based approach is being developed to estimate regional uptake from SPECT projections while accounting for intra-regional heterogeneity~\cite{li2024win}. 
Extending the proposed method to model intra-regional heterogeneity is an important area of future research.

Fig.~\ref{fig:mismatch} illustrates that inaccuracies in VOI definitions can affect the accuracy of the estimates produced by both the proposed and conventional RBQ methods. In these results, as the degree of inaccuracy in VOI definition due to rigid-body transformation increased, the accuracy of all the considered methods decreased. However, at each degree of inaccuracy, the proposed MEW-PDQ method consistently outperformed the conventional RBQ methods. Additionally, these results show that, compared with the conventional RBQ methods, the proposed method also exhibited lower sensitivity to inaccuracy in VOIs arising from non-rigid transformations caused by patient breathing during the SPECT scanning process.

For dosimetry studies with systematically administered radionuclide therapies, the patients are typically imaged at multiple time points post-administration. Due to the decay of $\mathrm{^{227}Th}$ to $\mathrm{^{223}Ra}$, there will be different $\mathrm{^{227}Th}$ to $\mathrm{^{223}Ra}$ uptake ratios in the patient at those time points. Fig.~\ref{fig:timeAfter} shows the reliable performance of the MEW-PDQ method to estimate the regional uptake of both isotopes with all $\mathrm{^{227}Th}$ to $\mathrm{^{223}Ra}$ uptake ratios. These results motivate further evaluation of the method for $\mathrm{^{227}Th}$-based $\alpha$-RPT dosimetry studies.

The MC-based approach generated system matrix of the proposed method allows for accurate modeling and compensation of crosstalk contamination and the other complicated SPECT physics in $\mathrm{^{227}Th}$-based $\alpha$-RPTs. This MC-based approach was relatively fast, and the system matrix can be pre-computed and stored given the small number of VOIs. For a typical simulated patient in this study, computing the system matrix of both isotopes for all VOIs took less than 10 minutes on our system equipped with Intel(R) Xeon(R) Gold 6230 CPU with 80 cores. The system matrix required only 120 MB for storage and memory during computation. 
In contrast, a system matrix generated for the DOSEM-based method using the same MC-based approach can be as large as 40 TB for a single patient, which is not feasible for processing and storage with typical computational systems.
Additionally, due to the low number of unknowns, the 1000 iterations of the MEW-PDQ method required less than 10 minutes on a desktop computer with a 16-core Intel Core i7-10700K CPU. Thus, the proposed method has the capability to efficiently perform system matrix modeling and regional uptake estimation with modest computational requirements.

The proposed method is designed to directly estimate uptake within volumes of interest, thus providing inputs for organ and lesion dosimetry, which is often the clinical task of interest~\cite{sgouros2020dosimetry,pandit2021dosimetry}. However, since the method does not yield a voxel-based reconstruction, it cannot be used for applications that require such reconstructions, such as voxel-based dosimetry. For these applications, there remains an important need to develop voxel-based reconstruction methods for $\mathrm{^{227}Th}$-based $\alpha$-RPTs, and this is an important area of future research. For example, if it is known that the VOIs do not have fully homogeneous uptake but have relatively homogeneous uptake in some VOIs and highly heterogeneous uptake in other VOIs, reconstruction methods could represent the VOIs using a small number of voxels with non-uniform sizes, similar to those proposed in previous studies~\cite{li2011adaptive,meng2009non}. Similar to the proposed method, such methods could make the estimation problem less ill-posed and reduce the noise in the estimated voxel values. 
Additionally, more recently, voxel-based reconstruction methods that use list-mode data and avoid binning-related information loss have been developed for single-isotope $\alpha$-RPTs~\cite{rahman2020list}.
Reconstructions from list mode data could extract the maximum possible information from the limited number of detected photons, thereby enhancing the quality of reconstructed images. 
Overall, developing and validating advanced voxel-based image reconstruction methods for $\mathrm{^{227}Th}$-based $\alpha$-RPTs is an important area of research; future studies could also compare the performance of the proposed method with RBQ methods that are based on such reconstructions.

A further limitation of this study is that the proposed method was evaluated in simulation studies. Although we simulated realistic patient phantoms, used MC software to simulate the SPECT system, and validated our SPECT system simulations with physical phantoms, we recognize that our simulations may not model all aspects of patient population variabilities and system instrumentation. Evaluation of the method with physical phantoms and patient data, including data from ongoing clinical trials such as NCT03724747, provides a mechanism to address these limitations. Additionally, we only simulated the GE Discovery 670 scanner with HEGP collimator. In clinics, patients may be imaged with various SPECT scanners and collimator configurations. Assessing the performance of the proposed method across different scanner-collimator configurations, especially the reproducibility of the estimates, is an important area for future research. 
We do note here that we have previously validated another projection-domain quantitative SPECT method for $\mathrm{^{223}Ra}$-based $\alpha$-RPTs across nine different scanner-collimator configurations, and observed that the method was accurate and highly reproducible across these different configurations~\cite{LiJNMaccepted}. This provides us with confidence that the proposed MEW-PDQ method would also yield reproducible findings, but this warrants further investigation. 
Also, due to the absence of reliable radiopharmacokinetic data, we modeled the regional uptake ratios of $\mathrm{^{227}Th}$ based on clinical studies using $\mathrm{^{223}Ra}$. Additionally, the biological half-life of these isotopes was not considered in the experiments that evaluated the performance of the proposed method at various $\mathrm{^{227}Th}$ to $\mathrm{^{223}Ra}$ uptake ratios. However, the application of the MEW-PDQ method in future clinical trials could contribute to reliably obtaining such radiopharmacokinetic data of $\mathrm{^{227}Th}$ and $\mathrm{^{223}Ra}$. 

\section{Conclusion}
We proposed a multiple-energy-window projection-domain quantitative SPECT (MEW-PDQ) method for joint regional uptake quantification of $\mathrm{^{227}Th}$ and $\mathrm{^{223}Ra}$. The method was observed to yield reliable (accurate and precise) estimates of the regional uptake of both isotopes, as evaluated using clinically realistic and validated simulation studies in the context of $\mathrm{^{227}Th}$-based $\alpha$-RPT, for multiple variations in lesion properties, including variations in lesion sizes, lesion-to-bone uptake ratios, and $\mathrm{^{227}Th}$~to~$\mathrm{^{223}Ra}$ uptake ratios, as well as in a virtual imaging trial. The proposed method consistently outperformed conventional reconstruction-based quantification approaches including a dual-isotope ordered subset expectation maximization (DOSEM)-based method and a geometric transfer matrix (GTM)-based method. We also observed that the proposed method yielded estimates that were almost unbiased and had a standard deviation close to that derived from the Cram\'er-rao lower bound, indicating that the method may be an efficient estimator. Moreover, we observe that the use of multiple energy windows can significantly increase the precision of the proposed method. Further, even with moderate levels of intra-regional uptake heterogeneity and inaccuracy in the region definitions, the proposed method still yielded reliable estimates and outperformed both the DOSEM and GTM-based methods. Overall, the results provide strong evidence supporting further evaluation and application of this method for performing quantitative SPECT of $\mathrm{^{227}Th}$-based $\alpha$-RPTs.

\section*{Acknowledgment}
All authors declare that they have no known conflicts of interest in terms of competing financial interests or personal relationships that could have an influence or are relevant to the work reported in this paper.

\setcounter{section}{0}
\renewcommand{\thesection}{S-\Roman{section}}
\begin{center}
\textbf{\huge Supplementary Materials}
\end{center}

\section{Validating the SPECT simulation}
\label{sec:validationSPECT}
\subsection{Experiments}
SIMIND is a Monte Carlo (MC) approach that has already been shown to model single-photon emission computed tomography (SPECT) imaging systems accurately~\cite{ljungberg1989monte,toossi2010simind,morphis2021validation}, including when imaging other $\alpha$-particle-emitting isotopes~\cite{li2022projection}. To demonstrate the accuracy of the SIMIND-based simulation approach for $\mathrm{^{227}Th}$-based $\alpha$-particle radiopharmaceutical therapies ($\alpha$-RPT) SPECT, we compared the projection data and energy spectra obtained with our simulation approach to that obtained on a physical scanner. 

To compare simulated and physically acquired projection data, we used a NEMA phantom (Data Spectrum$^\mathrm{TM}$, USA). The spheres of this phantom were filled with $\mathrm{^{227}Th}$ solutions with an activity concentration of 40~kBq/ml. The rest of the phantom was filled with water to simulate attenuation and scatter due to soft tissue. The phantom was scanned on a GE Discovery 670 SPECT/CT system with a medium energy general purpose (MEGP) collimator one day after purifying the $\mathrm{^{227}Th}$ isotope and filling the isotope in the phantom. Thus, at the time of scanning, a small portion of the $\mathrm{^{227}Th}$ had decayed to $\mathrm{^{223}Ra}$. During imaging, projections were acquired in two energy windows, corresponding to the two major photopeaks of $\mathrm{^{223}Ra}$ (66 - 98 keV) and $\mathrm{^{227}Th}$ (217 - 260 keV) at 60 angular positions spaced uniformly over 360$^{\circ}$. The same image-acquisition process was modeled using our simulation approach (Sec. III B in the main manuscript). In the simulation, the concentration of $\mathrm{^{227}Th}$ and $\mathrm{^{223}Ra}$ were calculated theoretically, based on the filling and scanning time. The profiles of the projection data obtained with the physical scanner and with the simulation approach from the two energy windows were compared. 

To demonstrate the accuracy of the SIMIND-based approach in simulating the emission spectra of $\mathrm{^{227}Th}$ and $\mathrm{^{223}Ra}$ and in simulating the energy resolution of the SPECT system, we compared the simulated and physically acquired energy spectra of each isotope. First, we measured the energy spectrum of each isotope on the GE Discovery 670 SPECT/CT system equipped with a high energy genera purpose (HEGP) collimator one day after purification. 
For imaging each isotope, we placed the source inside a 5 ml glass vial in the center of the field of view, a point 17 cm from both heads of the scanner. 
We acquired projections for 5 minutes at only two opposite angular positions using the dual-head system. Energy windows, as presented in Table~\ref{tab:energy_win_spectra}, were selected to provide coverage of the relevant photopeaks for each isotope. 
Next, we simulated the same image-acquisition process using our SIMIND-based approach (Sec. III B in the main manuscript). In this simulation, we assumed $\mathrm{^{223}Ra}$ and its daughters were at radioactive equilibrium. The activity of $\mathrm{^{227}Th}$ and $\mathrm{^{223}Ra}$ in each source was estimated based on the acquired projections.
We also simulated the stray-radiation-related noise in this process. To model the level of this noise, we estimated the mean value of stray-radiation-related noise in each projection bin for each energy window from the corresponding physically acquired projections. This was achieved by averaging the values of pixels close to the edge of the field of view. We chose these pixels because they were far from the vial and were expected to contain only stray-radiation-related noise.


\begin{table*}[]
\centering
\caption{Energy window settings for isotope spectra acquisition}
\label{tab:energy_win_spectra}
\resizebox{0.8\textwidth}{!}{%
\begin{tabular}{cccccccc}
\hline
Window   Index & Range (keV) & Window Index & Range (keV) & Window Index & Range (keV) & Window Index & Range (keV) \\ \hline
1 & 57.6   - 62.4 & 9  & 95.0   - 105.0  & 17 & 191.0   - 209.0 & 25 & 290.0   - 300.0 \\
2 & 62.5   - 67.5 & 10 & 105.1   - 115.1 & 18 & 210.1   - 229.9 & 26 & 300.1   - 310.1 \\
3 & 67.6   - 72.6 & 11 & 115.2   - 124.8 & 19 & 230.0   - 240.0 & 27 & 310.2   - 320.0 \\
4 & 72.7   - 77.5 & 12 & 124.9   - 133.8 & 20 & 240.1   - 249.9 & 28 & 320.1   - 330.1 \\
5 & 77.6   - 82.4 & 13 & 135.4   - 144.4 & 21 & 250.0   - 260.0 & 29 & 330.2   - 340.0 \\
6 & 82.5   - 87.5 & 14 & 144.8   - 155.2 & 22 & 260.1   - 270.1 & 30 & 340.7   - 379.9 \\
7 & 87.6   - 92.0 & 15 & 155.4   - 170.0 & 23 & 270.2   - 280.0 & 31 & 380.0   - 425.6 \\
8 & 92.1   - 94.9 & 16 & 170.5   - 189.5 & 24 & 280.1   - 289.1 & 32 & 425.7 - 474.7   \\ \hline
\end{tabular}%
}
\end{table*}

\subsection{Results}
\begin{figure*}
    \centering
    \includegraphics[width=0.7\textwidth]{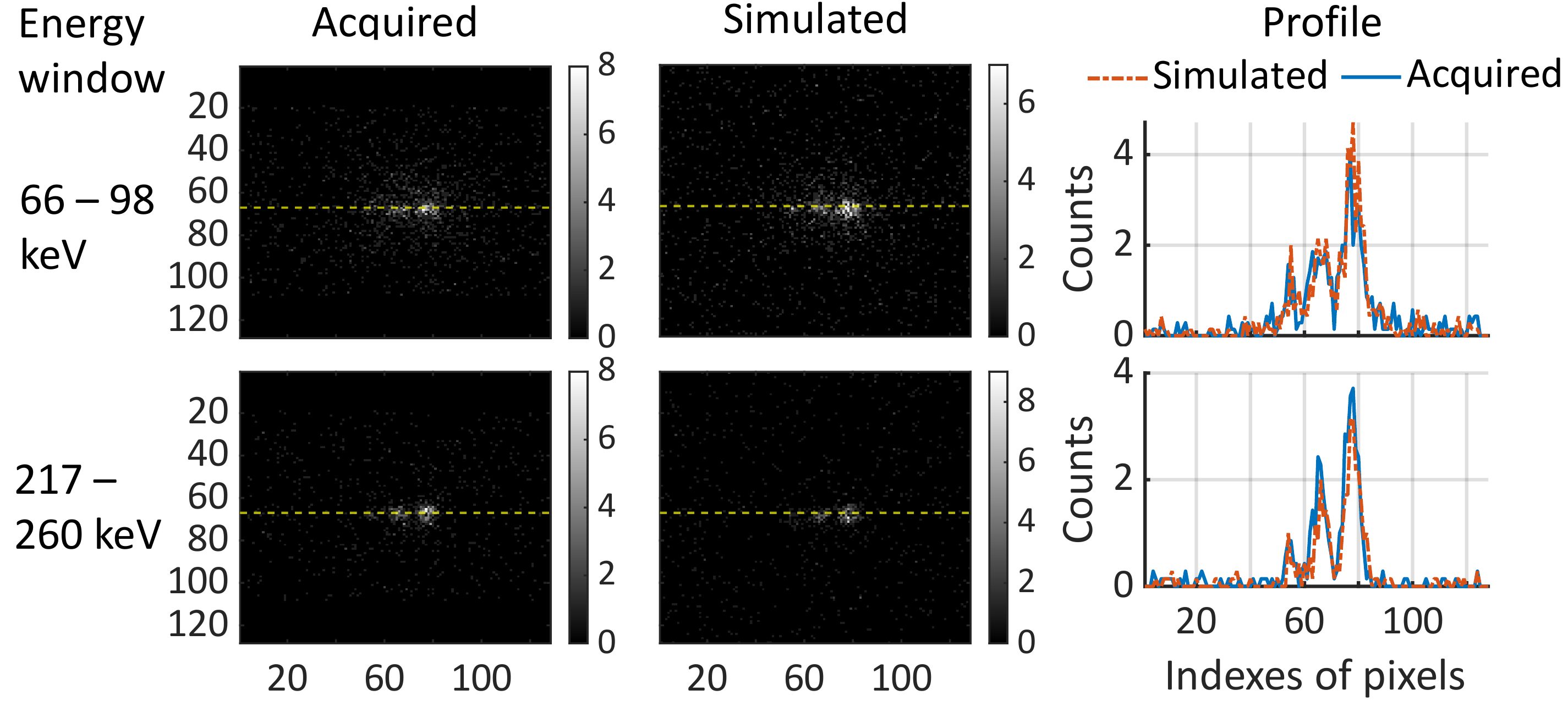}
    \caption{Comparison of the simulated and physical-SPECT-system-acquired
    projections of the NEMA phantom in two different energy windows.}
    \label{fig:realism_res}
\end{figure*}

Fig.~\ref{fig:realism_res} shows projections of the NEMA phantom in the two energy windows at the first angular position, acquired using simulated and physical SPECT systems. We also compared the profiles along the dashed line in the projections from the two approaches in both energy windows. To reduce the noise-related variation in this profile, each point in the profile was obtained by averaging the number of counts along five adjacent pixels on both sides of the dashed line. 
We observed that the profile of the simulated projection matched that acquired on the physical scanner in both energy windows. This provides evidence of the accuracy of the process to simulate isotope emission, the SPECT system, and the noise in this study.

\begin{figure*}
    \centering
    \includegraphics[width=0.8\textwidth]{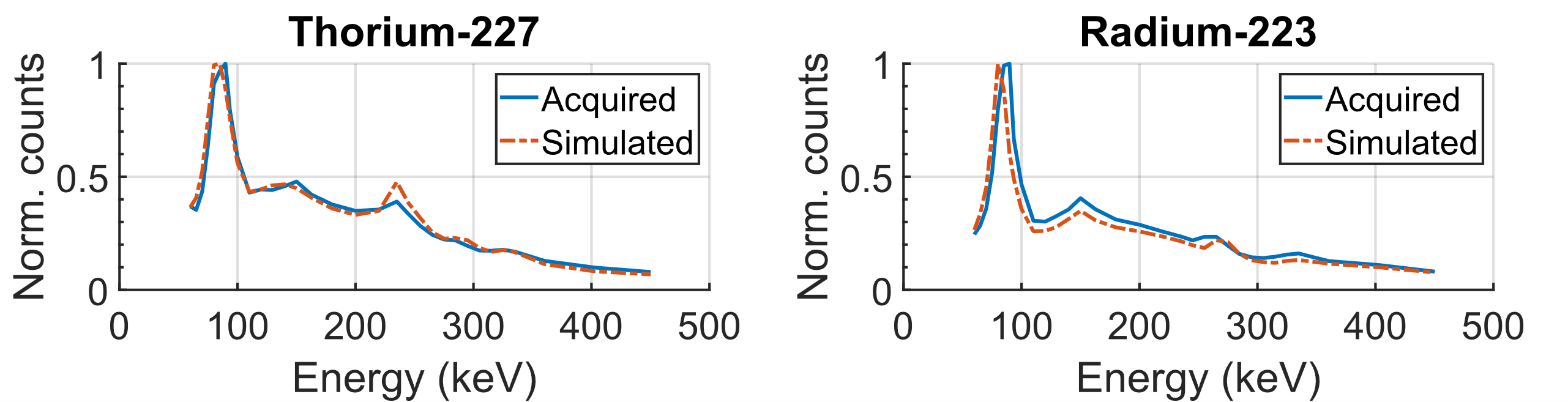}
    \caption{Comparison of the simulated and physical-SPECT-system-acquired spectra of $\mathrm{^{227}Th}$ and $\mathrm{^{223}Ra}$.}
    \label{fig:spectra_compare}
\end{figure*}

Fig.~\ref{fig:spectra_compare} presents both physically acquired and simulated spectra of $\mathrm{^{227}Th}$ and $\mathrm{^{223}Ra}$. 
The spectra were generated by dividing the total counts acquired in each energy window by the window width and plotting these values against the center of each energy window. Each spectrum has been normalized such that its peak value is unity. We observe that the simulated and physically acquired spectra matched for both isotopes. 
Those spectra were different from simply convoluting the emission spectra of the isotopes with the energy resolutions since there were significant amounts of stray-radiation-related noise and there was a finite number of energy windows. 

\section{Image reconstruction-based methods compared}
\subsection{Dual-isotope ordered subset expectation maximization image reconstruction (DOSEM)-based method}
Based on analysis of the emission spectra of both isotopes, energy windows 66 - 96 keV and 217 - 260 keV (energy windows 1 and 3 in Fig. 1 of the main manuscript) were considered as the primary energy windows of $\mathrm{^{223}Ra}$ and $\mathrm{^{227}Th}$, respectively.
We first reconstructed images of activity uptake of $\mathrm{^{227}Th}$ and $\mathrm{^{223}Ra}$ using projections from the corresponding two separate primary energy windows.
In this step, we assumed there was no crosstalk contamination, i.e., the counts in the two energy windows solely reflected emissions from $\mathrm{^{227}Th}$ and $\mathrm{^{223}Ra}$, respectively.
The images were reconstructed using the ordered subset expectation maximization (OSEM) method, implemented using Customizable and Advanced Software for Tomographic Reconstruction (CASToR) software~\cite{merlin2018castor}. In this process, we compensated for attenuation, scatter, collimator-detector response, stray-radiation-related noise, and the complicated emission spectra of both isotopes. The scatter was compensated using effective scatter source estimation (ESSE) method, in which the scatter kernels were generated using SIMIND simulations~\cite{frey1996new}.
Next, the reconstructed image of each isotope was forward projected to the photopeak energy window of the other isotope to estimate the crosstalk contamination in each photopeak energy window.   
In this forward projection, we modeled the emission spectra of both isotopes, attenuation, scatter, and collimator-detector response. In this step, the crosstalk contamination from primary photons was modeled from the emission spectrum, and the crosstalk contamination from scattered photons was modeled with the ESSE method. Then, the images of both isotopes were again reconstructed using the OSEM method, with the estimated crosstalk contamination as additive correction terms. Since the estimates of crosstalk contamination were not accurate due to the inaccurate reconstruction in the first step, we repeated these steps for multiple iterations. 

More specifically, we denote the process of reconstructing and forward projecting the images of both isotopes to compensate for crosstalk as a 'joint iteration'. The iterations executed specifically for each isotope in the OSEM reconstruction are referred to as 'isotope-specific iterations'. For the purpose of fine-tuning the DOSEM-based method, we sought the optimal number of joint iterations and isotope-specific iterations. To this end, we generated 50 noise realizations of a representative patient phantom. Initially, image of each isotope was reconstructed from its corresponding photopeak energy window using the OSEM-based method, incorporating ideal crosstalk corrections. Through this process, we determined that the optimal number of isotope-specific iterations was 20, each comprising 6 subsets. With this number of isotope-specific iterations, the normalized root mean square error (NRMSE) between the true and estimated lesion uptake from the reconstructed images of both isotopes converged to its lowest value. Convergence was defined as a change in NRMSE of less than 1\% over the next 10 iterations. Subsequently, we performed the DOSEM-based reconstruction from the crosstalk-contaminated SPECT projections as outlined above. The optimal number of joint iterations was established at 5, ensuring that changes in the mean estimated uptake of each isotope in each region were less than 0.1\% with additional joint iterations.

\subsection{Geometric transfer matrix (GTM)-based method}
Partial volume effects (PVEs) are known to degrade quantification accuracy in SPECT~\cite{soret2007partial}. The proposed method implicitly assumes constant uptake within each volume of interest (VOI). Under this assumption, PVEs can also be compensated for post-reconstruction.
We compared our approach to the widely used geometric transfer matrix (GTM)-based method~\cite{rousset1998correction}. For each isotope, the elements of GTM were calculated from the reconstructed projections of the VOIs, obtained as described in Sec. III A of the main manuscript.
Additional implementation details of this method are described in~\cite{rousset1998correction}.

\section{Figures of merit}
For a range of experimental conditions, we generated multiple instances of projection data for a single phantom, where each instance corresponded to separate noise realizations.
Denote the total number of realizations by $R$. Considering one of the isotopes, denote the true activity uptake of the $k^{th}$ VOI by $\lambda_{k}$  and the corresponding estimate with the $r^{th}$ noise realization by $\hat{\lambda}_{rk}$. In these experiments, the accuracy of the estimated uptake of this isotope was quantified using the normalized bias (NB), which, for the $k^{th}$ VOI, is given by
\begin{equation}
\label{eq:NB}
    \mathrm{NB}_{k}=\frac{1}{R}
    \sum\limits_{r=1}^{R} \frac{\hat{{\lambda }}_{rk}-{{\lambda }_{k}}}{{{\lambda }_{k}}}.
\end{equation}
The precision of the estimated uptake of this isotope was quantified using the normalized standard deviation (NSD), which, for the $k^{th}$ VOI, is given by
\begin{equation}
\label{eq:NSD}
    \mathrm{NSD}_{k} = \sqrt{
    \frac{1}{R-1}
    \sum\limits_{r=1}^{R}{\left(\frac{\hat{{\lambda }}_{rk}}{\lambda_{k}}-\frac{1}{R}\sum\limits_{r'=1}^{R}\frac{\hat{{\lambda }}_{r'k}}{\lambda_{k}}\right)^{2}}
    }.
\end{equation}
Finally, the overall error in estimating the uptake was quantified by the normalized root mean square error (NRMSE). For the $k^{th}$ VOI,
\begin{equation}
    \mathrm{NRMSE}_{k} = \sqrt{NB_k^2+NSD_k^2
    }.
\end{equation}

Multiple experiments in this study were conducted across a plural number of patients. 
To evaluate the performance of the methods over a population of patients, each with one or multiple noise realizations, we used the ensemble NB and ensemble NRMSE.
We denote the number of samples in the population by $S$ and the number of noise realizations for each patient sample by $R$, and denote the true and estimated activity uptake of a particular isotope in the $k^{\th}$ VOI for the $s^{\th}$ sample from the $r^{th}$ realization by $\lambda_{sk}$ and $\hat{\lambda}_{skr}$, respectively. 
The ensemble NB for the $k^{th}$ VOI is given by
\begin{equation}
    \mathrm{Ensemble~NB}_{k}=\frac{1}{SR}
    \sum\limits_{s=1}^{S} \sum\limits_{r=1}^{R}
    \frac{\hat{{\lambda }}_{skr}-{{\lambda }_{sk}}}{{{\lambda }_{sk}}}.
\end{equation}
The ensemble NRMSE for the $k^{th}$ VOI is given by
\begin{equation}
    \mathrm{Ensemble~NRMSE}_{k} = \sqrt{\frac{1}{SR}
    \sum\limits_{s=1}^{S} \sum\limits_{r=1}^{R}
    \left(\frac{\hat{{\lambda }}_{skr}-{{\lambda }_{sk}}}{{{\lambda }_{sk}}}\right)^2}.
\end{equation}

Additionally, to quantify performance in cases where we had just a single estimate, we used normalized error, defined as the difference between the estimated and true uptake values, normalized by the true uptake value.

Finally, we also computed the Cram\'er-Rao lower bound (CRLB), which is the minimum variance that can be achieved by an unbiased estimator, as a benchmark for the precision of the activity estimated using the proposed method.
The CRLB is given by the diagonal elements of the inverse of the Fisher information matrix for the estimated parameter. We denote the Fisher information matrix by $\bm{F}$~\cite{barrett2013foundations}.
Since the task is to estimate regional uptake of $\mathrm{^{227}Th}$ and $\mathrm{^{223}Ra}$ at the same time, the Fisher information matrix needs to consider both isotopes. We denoted $\lambda_{l}$ and $\hat{\lambda}_{l}$ as the true and estimated activity uptake respectively, where $l$ ranges over the two isotopes in $K$ VOIs. Thus the Fisher information matrix is a $2K$ by $2K$ matrix with elements given by
\begin{equation}
    F_{l_{1}l_{2}}=-E\left[\frac{\partial^{2} }
    {\partial \lambda_{l_{1}} \partial \lambda_{l_{2}} }
    \ln \Pr (\bm{g}|\bm{\lambda})\right],
    \label{eq:CRLB1}
\end{equation}
where $E[x]$ denotes the expectation of a random variable $x$. 
We have already derived the likelihood of the measured data $\g$ in the main manuscript:
\begin{equation}
\begin{split}
    \Pr(\g|\bm{\lambda}) & = \prod_{m=1}^{M}\exp[-(\bm{H} \bm{\lambda})_m-\psi_m] \frac{[(\bm{H} \bm{\lambda})_m+\psi_m]^{g_m}}{{g_m}!}.
\end{split}
    \label{eq:poisson}
\end{equation}

Substituting Eq.~(\ref{eq:poisson}) in Eq.~\eqref{eq:CRLB1} yields
\begin{equation}
    F_{l_{1}l_{2}}=
    \sum\limits_{m=1}^{M}{\frac
    {H_{ml_1} H_{ml_2}}
    {
    (\bm{H} \bm{\lambda})_{m} + \psi_m
    }
    }.
    \label{eq:CRLB}
\end{equation}

\section{Performance of the LC-QSPECT in $\mathrm{^{227}Th}$-based $\alpha$-RPT}
\begin{table*}[t]
\centering
\caption{The performance of the MEW-PDQ and LC-QSPECT methods on the task of quantifying regional uptake of $\mathrm{^{227}Th}$ and $\mathrm{^{223}Ra}$ in $\mathrm{^{227}Th}$-based $\alpha$-RPTs.}
\label{tb:LCQ}
\begin{tabular}{ccccccccc}
            & \multicolumn{4}{c}{$\mathrm{^{223}Ra}$}       & \multicolumn{4}{c}{$\mathrm{^{227}Th}$}    \\ \hline
NB (\%)     & Background & Bone  & Gut   & Lesion  & Background & Bone & Gut  & Lesion  \\
MEW-PDQ     & -0.03      & -1.2  & 0.05  & 0.8     & 0.01       & 1.1  & -0.3 & 0.4     \\
LC-QSPECT   & 250.8      & 178.8 & 154.4 & 450.0   & 6.3        & 2.9  & 7.9  & 1.8     \\ \hline
NRMSE  (\%) & Background & Bone  & Gut   & Lesion & Background & Bone & Gut  & Lesion \\
MEW-PDQ     & 2.0        & 9.5   & 1.1   & 18.9    & 0.9        & 4.7  & 0.8  & 3.6     \\
LC-QSPECT   & 250.8      & 179.0 & 154.4 & 450.3   & 6.4        & 5.2  & 7.9  & 3.9    
\end{tabular}
\end{table*}

\subsection{Experiments}
In our previous studies~\cite{li2022projection}, we proposed the low-count quantitative SPECT (LC-QSPECT) method to estimate the regional uptake of a single isotope from its photopeak energy window projections. The method has been observed to accurately and precisely estimate the regional uptake for patients administered with \(\mathrm{^{223}Ra}\)-based $\alpha$-RPTs. Therefore, in this experiment, we evaluate the performance of the LC-QSPECT method on quantifying the regional uptake of $\mathrm{^{227}Th}$ and $\mathrm{^{223}Ra}$ for patients undergoing $\mathrm{^{227}Th}$-based $\alpha$-RPT.

For this purpose, as described in Sec.~III C of the main manuscript, we considered a patient phantom with average patient size, a 33.75 mm diameter lesion in the pelvis, standard total uptake, and regional uptake ratios as presented in Table 1 of the main manuscript. As described in Sec.~III B of the main manuscript, we generated 50 noise realizations of that patient. We estimated the regional uptake of $\mathrm{^{227}Th}$ and $\mathrm{^{223}Ra}$ from those noise realizations using both the LC-QSPECT and the proposed multiple-energy-window projection-domain quantification (MEW-PDQ) methods. More specifically, the LC-QSPECT method was applied to individually estimate the regional uptake of \(\mathrm{^{223}Ra}\) and \(\mathrm{^{227}Th}\) from their respective photopeak energy windows, without accounting for crosstalk contamination. More implementation details of the LC-QSPECT method are described in~\cite{li2022projection}. The MEW-PDQ method was applied as described in Sec.~III A of the main manuscript.
\subsection{Results}
Table~\ref{tb:LCQ} presents the NB and NRMSE values for the estimated regional uptake of \(\mathrm{^{227}Th}\) and \(\mathrm{^{223}Ra}\) obtained in this experiment. It was observed that the LC-QSPECT method yielded NB and NRMSE values exceeding 100\% in the estimation of \(\mathrm{^{223}Ra}\) uptake across all regions. Additionally, the proposed MEW-PDQ method significantly outperformed the LC-QSPECT method.

\section{CRLB with different energy window combinations}
\subsection{Experiments}
To supplement the evaluations in Sec. III C 1) d) of the main manuscript, we conducted an additional analysis to assess the impact of using more energy window combinations on the theoretical lower bound of estimated regional uptake variance. For this purpose, we calculated the CRLB for all possible combinations of the four energy windows depicted in Fig. 1 of the main manuscript. 
The same typical patient phantom was considered in this experiment and the CRLB-derived NSD values were calculated as described in Section III C 1) d) of the main manuscript.

\subsection{Results}
\begin{figure*}
    \centering
    \includegraphics[width=1.0\textwidth]{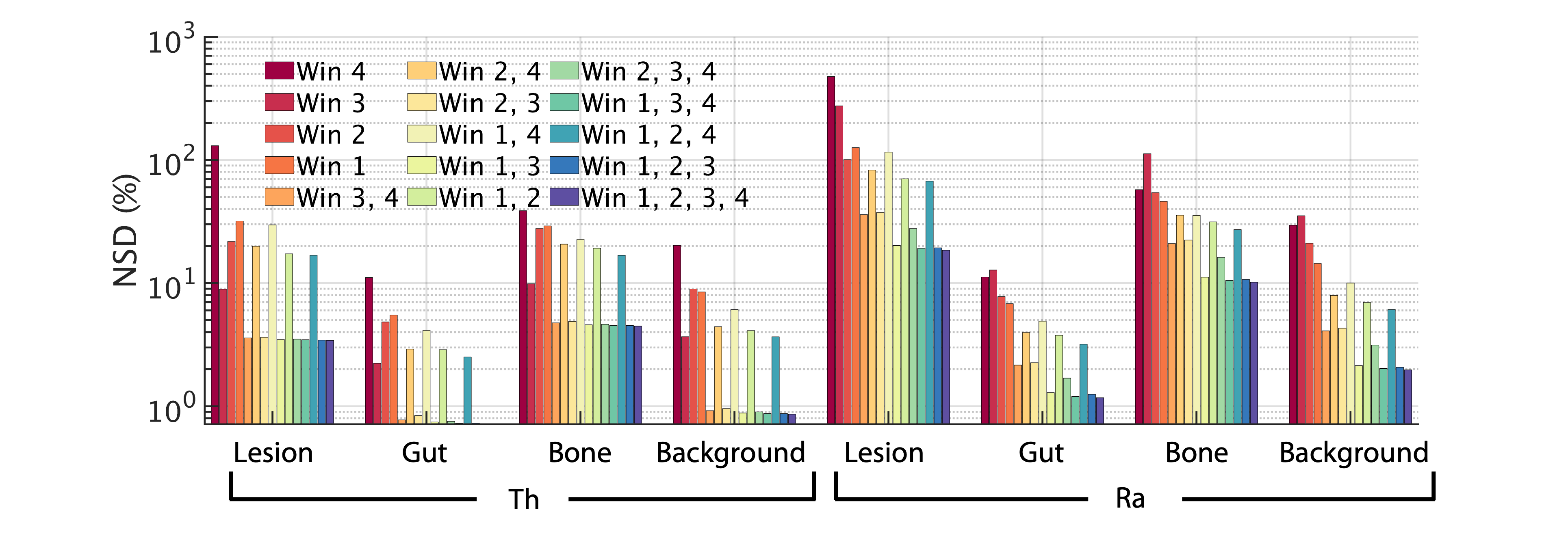}
    \caption{CRLB-derived NSD values for estimating regional uptake of $\mathrm{^{227}Th}$ and $\mathrm{^{223}Ra}$ using different combinations of energy windows.}
    \label{fig:CRLB_comb}
\end{figure*}

Fig.~\ref{fig:CRLB_comb} shows the CRLB-derived NSD values for estimating regional uptake of $\mathrm{^{227}Th}$ and $\mathrm{^{223}Ra}$ using different combinations of energy windows. We observe that using all four energy windows resulted in the lowest CRLB-derived NSD value, consistent with the data processing inequality principle~\cite{beaudry2011intuitive}. 

\bibliographystyle{IEEEtran} 
\bibliography{IEEEabrv,MEW_PDQ}
\end{document}